\documentclass[aps,reprint,prb,amsmath,amssymb]{revtex4-1}

\usepackage{graphicx}% Include figure files
\usepackage{dcolumn}% Align table columns on decimal point
\usepackage{bm}% bold math
\usepackage{subfig}
\usepackage[utf8]{inputenc}
\usepackage{amsmath,amssymb}
\usepackage{epstopdf}
\usepackage{textcomp}
\usepackage{hyperref}
\usepackage{listings}
\usepackage{upgreek}
\usepackage{xcolor}

\newcommand{\Fref}[1]{Fig.~\ref{#1}}
\newcommand{\Eqref}[1]{Eq.~(\ref{#1})}
\newcommand{\be}{\begin{equation}}
\newcommand{\ee}{\end{equation}}
\newcommand{\bal}{\begin{align}}
\newcommand{\eal}{\end{align}}
\newcommand{\bear}{\begin{eqnarray}}
\newcommand{\eear}{\end{eqnarray}}
\newcommand{\nn}{\nonumber}

\newcommand{\dm}{\mathrm{d}}
\newcommand{\e}{\mathrm{e}}
\newcommand{\cw}{C_\mathrm{W}}

\begin{document}

\title{Sound Absorption in Partially Ionized Hydrogen Plasma\\
and Heating Mechanism of Solar Chromosphere}

\author{Todor~M.~Mishonov}
\email[E-mail: ]{mishonov@bgphysics.eu}
\affiliation{Institute of Solid State Physics, Bulgarian Academy of Sciences,\\
72 Tzarigradsko Chaussee Blvd., BG-1784 Sofia, Bulgaria}
\affiliation{Faculty of Physics, 
St.~Clement of Ohrid University at Sofia,
5 James Bourchier Blvd., BG-1164 Sofia, Bulgaria}

\author{Iglika~M.~Dimitrova}
\affiliation{Faculty of Chemical Technologies,
University of Chemical Technology and Metallurgy,\\
8 Kliment Ohridski Blvd., BG-1756 Sofia}

\author{Albert~M.~Varonov}
\email[E-mail: ]{varonov@issp.bas.bg}
\affiliation{Institute of Solid State Physics, Bulgarian Academy of Sciences,\\
72 Tzarigradsko Chaussee Blvd., BG-1784 Sofia, Bulgaria}

\date{28 October 2020, 14:14}

\begin{abstract}
The temperature dependence of rates of electron impact
ionization and two electrons recombination are calculated
using Wannier cross section of electron impact ionization
of neutral hydrogen atom.
Entropy production and power dissipation are derived
for the case when the ionization degree deviates from its equilibrium value.
This is the special case of the obtained general formula for entropy production
accompanying chemical reactions.
Damping rate of the sound waves is calculated 
and the conditions when ionization processes dominate are considered.
A quasi-classical  approximation for the heating mechanism of solar 
chromosphere is proposed.
Several analogous phenomena for damping rates in liquids and crystals are shortly discussed,
for example, deaf sound of a glass of beer or English salt solution.
An explicit expression for the second or bulk (or volume) viscosity of hydrogen plasma is 
calculated from firsts principles. 
For the first time some second viscosity is calculated from first principles.
\end{abstract}

\maketitle

%%%%%%%%%%%%
\section{Introduction}
In many cases the absorption of sound waves cannot be described by hydrodynamic equations without taking into account the dispersion, i.e. the frequency dependence of the second viscosity.
The oscillations of the pressure of the longitudinal sound 
waves create oscillations of the temperature.
Further temperature oscillations change the 
equilibrium densities participating in the chemical 
equilibrium constants.
In such a way the propagation of sound waves
induces small oscillations of the chemical composition of the medium.
The forced chemical reactions and oscillations of the compound however are related with average increase 
of the entropy at almost constant temperature.
This increase of the entropy gives an 
irreversible energy dissipation and this dissipation 
creates an additional damping rate of the sound waves.

Within the hydrodynamic approach, the sound absorption 
is proportional to the square of the frequency $\omega^2$,
while the new mechanism proposed in the present 
article gives a constant damping rate.
Definitely at some low frequencies the considered mechanism dominates.

The hydrogen gas is actually a very simple object 
and in this model case all steps in the listed above chain of reasoning can be performed  analytically.
Starting from the classical  consideration by Wannier\cite{Wannier:53}
on near threshold electron impact ionization of the hydrogen (H) atom in ground state we calculate analytically the chemical rates of electron ionization and corresponding to the
inverse process of two electron recombination.
Further, we derive the kinetic equations for the 
densities of electrons $n_e$, protons $n_p$,
neutral atoms $n_0$, time dependence of the temperature
$T(t)$ and the time derivative of entropy which is one of the central results.
The energy dissipation related to entropy production gives 
the explicit expression for the extinction of sound waves
and all details of the derived formula can be calculated from first principles and compared with the experiment;\cite{Fite:58,Rothe:62,McGowan:67,Shyn:92}
for a contemporary review see Ref.~\onlinecite{Shakhatov:18}.
In the quasi-classical  approximation the derived sound absorption
reveals the mechanism of heating of hydrogen plasma
in the partially ionized solar chromosphere.
Physics of some related phenomena in liquids and crystals 
are shortly analyzed.

The paper is organized as follows:
in the next section~\ref{Hydrogen atom ionization by electron impact}
hydrogen atom ionization by electron impact is considered,
then in Sec.~\ref{Electron density kinetics}
electron density kinetics is analyzed.
The kinetic equation for the entropy
is derived in Sec.~\ref{Kinetic equation for the entropy}.
The derived result are applied for the
calculation of sound absorption in
Sec.~\ref{Sound absorption}.
Finally in Sec.~\ref{M-L}
the Mandelstam-Leontovich theory for the influence of slow chemical reactions on sound propagation is applied.
This theory published in 1937 and
from the same year are the first precise measurements of the sound absorption 
(supported by the navy, one can easily check for citations of Leontovich-Mandelstam-Landau-Lifshitz 
theory in the contemporary navy researches on the influence of MgSO$_4$
solvatation on sound damping in see water) 
in the oceans remained classified for 50 years. 
In order Tom, Dick and Harry to use Mandelstam-Leontovich theory,
Landau and Lifshitz re-derived their results in a transparent way.

The same words can be said about the sound propagation in partially
ionized hydrogen. 
Hydrogen projects related to nuclear synthesis completely
inhibited the regular research in plasma kinetics.
We will be not surprised if the results of the present work are actually 
well known and developed simultaneously and independently by colleagues
working on hydrogen projects in US (+SR).
After the elimination of such intellectual efforts university science reaches the same level
half a century later.
Now the launching of the Parker Solar probe triggers the opening of the Pandora box
of the kinetics of partially ionized hydrogen plasma and
in general the use of sound damping for monitoring of chemical reactions in fluids.

%%%%%%%%%%%%%%%%%%%%%%%%%%%%
\section{Hydrogen atom ionization by electron impact}
   \label{Hydrogen atom ionization by electron impact}
   
For electron energies $\varepsilon$ slightly above the ionization threshold $I$ correlated motion of two electrons
in the final state is almost classical and for this difficult 
problem Wannier obtained the well-known 
result for the ionization cross-section\cite{Wannier:53,LL3} 
\begin{align}
& \sigma(\varepsilon) = \cw a_\mathrm{B}^2 
\left ( \frac{\varepsilon}{I} -1 \right )^{\! w},
\end{align}
where $w\approx 1.18$, 
$\cw$ is a dimensionless constant and 
\begin{align}
a_\mathrm{B}
=\frac{\hbar c}{e^2}\,\frac{\hbar}{mc}
=\frac{\hbar^2}{me^2},\quad
e^2\equiv\frac{q_e^2}{4\pi\varepsilon_0},
\quad
\frac{e^2}{\hbar c}\approx\frac1{137}
\end{align}
is the Bohr radius.
\textcolor{black}
{
We use SI units in which the electron charge 
$q_e=1.602\times 10^{-19}\,\mathrm{C}$
but all formulae in the present paper are given in system invariant form.}

For the rate of the ionization reaction
$\mathrm{H} + \mathrm{e} 
\longrightarrow \mathrm{p} + \mathrm{e} + \mathrm{e}$
\begin{align}
\beta = \langle v \sigma(\varepsilon) \rangle, 
\qquad  v = \sqrt{2\varepsilon/m},
\end{align}
the integration on the Maxwell velocity $v$ distribution
\begin{align}
&
\label{beta}
\beta = 
2 \sqrt{\frac{2}{\pi}} \cw a_\mathrm{B}^2  v_{Te} 
\int\limits_\iota^\infty  ( \epsilon - \iota  )^w  
\, \epsilon^{1/2} \, \e^{-\epsilon} \dm \epsilon, \\
&
v_{Te} \equiv \sqrt{T/m},\qquad
\epsilon \equiv \varepsilon/T ,\qquad
\iota \equiv I/T\gg1
\label{iota}
\end{align}
for low temperatures $T\ll I$ gives
\begin{align}&
\beta\approx C_\beta \beta_\mathrm{B}\,\e^{-\iota},
\quad
C_\beta
\equiv 2\Gamma(w+1) C_\mathrm{W}/\sqrt{\pi},
\label{ionization_rate}
\\&
\beta_\mathrm{B}\equiv v_\mathrm{B}a_\mathrm{B}^2,
\quad
v_\mathrm{B} =(e^2/\hbar c)\, c \equiv e^2/\hbar;
\end{align}
\textcolor{black}
{temperature is given in energy units $T=k_\mathrm{B}T^\prime$,
the Boltzmann constant
$k_\mathrm{B}$ time temperature in Kelvins $T^\prime$~[K].}
In other words the ionization potential $I$
is the activation energy of the rate or the chemical reaction
parameterized by the Bohr velocity $v_\mathrm{B}$ and natural 
for the atomic physics rate unit 
$v_\mathrm{B}a_\mathrm{B}^2$.
According to the experimental study by McGowan and Clarke\cite{McGowan:67} 
$w \approx 1$ and the slope of the almost linear dependence
depicted in their Fig.~6 gives $C_\mathrm{W}\approx 2.7$
whence according to Eq.~(\ref{ionization_rate})
we calculate $C_\beta\approx 3.0\simeq 1.$

The considered process gives for the time derivative
of the electron density 
\begin{align}
\dm n_e /\dm t \equiv \dot n_e \approx \beta n_0  n_e,
\end{align}
where $n_0$ is the volume density of neutral atoms.
In the next section we consider other processes 
in the hydrogen plasmas.

%%%%%%%%%%%%%%%%%%
\section{Electron density kinetics}
   \label{Electron density kinetics}
In the general case we have to take into account
the rate of many processes 
\begin{align}
\beta_\varphi: \quad & \mathrm{H} +\gamma  \longrightarrow \mathrm{p} + \mathrm{e}, \\
\beta: \quad & \mathrm{H} + \mathrm{e} \longrightarrow \mathrm{p} + \mathrm{e} + \mathrm{e}, \\
\beta_{_\mathrm{P}}: \quad & \mathrm{H} + \mathrm{H} \longrightarrow \mathrm{p} + \mathrm{e}+\mathrm{H}, \\
\gamma_\varphi: \quad & \mathrm{p} + \mathrm{e} \longrightarrow  \mathrm{H} +\gamma, \\
\gamma: \quad & \mathrm{p} +\mathrm{e} + \mathrm{e} \longrightarrow \mathrm{H} + \mathrm{e},\\
\gamma_{_\mathrm{P}}:\quad &
(\mathrm{p} + \mathrm{e})^*+\mathrm{H} \longrightarrow
\mathrm{H}+\mathrm{H},
\end{align}
and more complete set of kinetic equations
for the densities of electrons $n_e$, protons $n_p$
and neutral atoms $n_0$ looks like
 \begin{align}&
\dot n_e=\dot n_e^{(+)}-\dot n_e^{(-)},
\qquad \dot n_e=\dot n_p=-\dot n_0,
\label{kinetic_equation}
\\&
\dot n_e^{(+)}=\beta_\varphi n_0 
+ \beta n_0 n_e
+ \beta_{_\mathrm{P}} n_0^2 ,\\&
\dot n_e^{(-)}
=\gamma_\varphi n_p n_e 
+ \gamma n_p n_e^2
+ \gamma_\mathrm{_P}n_en_pn_0.
\end{align}

Let us gives for reference the rate 
of direct process of radiation recombination of the free
electrons directly to the ground state of the H atom
\begin{align}
&
\gamma_\varphi=C_\varphi
\left(\frac{\hbar}{mc}\right)^{\!\! 3}\,
       \frac{I}{\hbar}\,\sqrt{\iota},
\quad 
C_\varphi=\frac{2^{10} \pi^3}{6\e^4}\approx 34.8,
\\&
\sigma_\mathrm{\varphi}=C_\varphi
\frac{(e^2/\hbar c)a_\mathrm{B}^2I^2}{m^2c^2v^2}. 
\end{align}
The principle of the detailed balance is used 
in the derivation of the rate of the recombination $\gamma_\varphi$ 
using the low energy cross-section
of photo-ionization $\sigma_\mathrm{\varphi}$.\cite{LL4}

Another process is the neutral atom catalyzed 
rate of radiationless recombination
at low temperatures considered by 
Pitaevskii\cite{Pitaevskii:62,LL10} 
\begin{align}
& \gamma_\mathrm{_P}=C_\mathrm{_P}
 \frac{e^6 \sigma_{e0}}{MT^2}  \sqrt{\frac{m}{T}},\\
&C_\mathrm{_P}=\frac13 32 \sqrt{2 \pi} \approx 26.7,
\qquad T\ll I\sqrt\frac{m}{M},
\end{align}
where $M$ is the proton mass;
see also Ref.~\onlinecite{BelyaevBudker:58} and references
therein.

The rates $\bar n_p \bar n_e\gamma_\varphi
=\bar n_0 \beta_\varphi$ are related by the principle
of the detailed balance\cite{LL3,LL10}
followed by thermal averaging
\be
\frac{\bar n_p \bar n_e}{\bar n_0 }
=\frac{\beta}{\gamma}
=\frac{\beta_\varphi}{\gamma_\varphi}
=\frac{\beta_{_\mathrm{P}}}{\gamma_{_\mathrm{P}}},
\label{balance}
\ee
where $\bar n_e$, $\bar n_p$, and $\bar n_0$
are the equilibrium densities for electrons, protons and neutral atoms.

In order to recall the equations for thermal equilibrium 
following\cite{Hill}
let us introduce convenient notations for the partition functions \cite{KittelCramer, Hill}
\begin{align}
&
n_Q\equiv \left ( \frac{MT}{2 \pi \hbar^2} \right )^{3/2}, \qquad
n_q \equiv \left ( \frac{mT}{2 \pi \hbar^2} \right )^{3/2},
\label{Saha}\\
&
n_{_\mathrm{S}}\equiv n_q\e^{-\iota}
\equiv \left ( \frac{mT}{2 \pi \hbar^2} \right )^{3/2}
\e^{-I/T}, \quad \frac1{K_p(T)}\equiv
p_{_\mathrm{S}}\equiv n_\mathrm{_S} T \nn. 
\end{align}
In this case the chemical balance of the reaction $\mathrm{H}\leftrightarrow \mathrm{p} + \mathrm{e}$
reads\cite{LL5,KittelCramer,Hill}	
\be
\frac{\bar n_p \bar n_e}{\bar n_0 }
=n_{_\mathrm{S}}(T).
\label{Saha_2}
\ee
In our approach the degree of ionization $\alpha$\cite{LL5,Saha}
is an \textit{additional state variable}
\begin{align}&
\alpha\equiv\frac{n_p}{n_\rho},\quad
n_\rho=n_0+n_p=\rho/M,\quad
n_\mathrm{tot}=n_0+n_p+n_e,\nn\\&
n_e=n_p=\alpha \,n_\rho,\quad n_0=(1-\alpha) \,n_\rho,
\quad n_\mathrm{tot}=(1+\alpha)n_\rho.\nn 
\end{align}
The essence of the Mandelstam and Leontovitch\cite{Mandelstam:37} theory
is to introduce slowly relaxing state variable denoted in the Landau-Lifshitz\cite{LL6}
course by $\xi$ and later on called by De Groot and Mazur\cite{DeGroot}
\textit{degree of advancement}. 
In some sense $\xi$ is analogous to the order parameter $\tilde\eta$
in Landau (1937) theory of type-II phase transitions.\cite{LL5}
For our problem the pressure is also expressed by $\alpha$
\be
p(T,\rho,\alpha)=n_\mathrm{tot}T=(1+\alpha)n_\rho T,
\quad
\rho=Mn_\rho.
\label{pressure}
\ee
can be expressed by the mass density of the gas
$\rho=Mn_\rho$ or by the pressure $p=nT$,
in the textbook Ref.~\onlinecite{Hill} 
numbers of particles per unit volume $n_e$, $n_p$ and $n_0$ 
are denoted as 
$\rho_{e^-}$, $\rho_{\mathrm{H}^+}$ and $\rho_{\mathrm{H}}$.
The substitution of these $n_e,$ $n_p$ and $n_0$
in Eq.~(\ref{Saha_2}) gives\cite{LL5}
\begin{align}
\label{p_alpha_n}
\frac{p}{p_\mathrm{_S}}
&=\frac{1-\bar\alpha^2}{\bar\alpha^2}
=\frac1{\bar\alpha^2}-1,\qquad\quad
\dfrac{n_\rho}{n_\mathrm{_S}}
=\dfrac{1-\bar\alpha}{\bar\alpha^2},
\\
\bar\alpha&=\frac1{\sqrt{1+K_p(T)p}}\label{alpha_Saha}\\
&= \dfrac{1}{\sqrt{1+\dfrac{p}{p_{_\mathrm{S}}(T)}}}
=\frac{-1+\sqrt{4\,\dfrac{n_\rho}
{n_\mathrm{_S}(T)}+1}}
{2\,\dfrac{n_\rho}{n_\mathrm{_S}(T)}}\in (0,1).
\label{bar_alpha}
\end{align}
% where $g_0/g_pg_e=1$ is taken into account.
For completeness we give the standard representation of the Saha equation
and its solution applied for hydrogen plasma\cite{LL5,Hill} 
which are expressed by dimensionless concentrations  
\be
\frac{\mathcal{C}_0}{\mathcal{C}_e\tilde{\mathcal{C}}_p}=pK_p(T),
\ee
\be
\mathcal{C}_e\equiv\frac{n_e}{n}=\mathcal{\tilde{C}}_p\equiv\frac{n_p}{n}=\frac{\alpha}{1+\alpha},\quad
\mathcal{C}_0=\frac{n_0}{n}=\frac{1-\alpha}{1+\alpha}.
\label{n_alpha}
\ee
Using the so introduced notations for plasma with constant density 
$n_\rho=\mathrm{const}$ we obtain the dynamic equation 
for the degree of ionization
\begin{align}
&
\frac{\mathrm{d}\alpha}{\mathrm{d}\theta}
=-\left[(1-\overline{\alpha})\alpha^3-\overline{\alpha}^2
(1-\alpha)\alpha\right]/\overline{\alpha}^2,
\label{dot_alpha}
\\
&
\theta \equiv
\int\limits_0^t\tilde\nu_\beta(t^\prime)\,\mathrm{d}t^\prime
\approx
\tilde\nu_\beta t, \quad \tilde\nu_\beta=\beta n_\rho,\\
&
\frac{\overline{\alpha}+(1-\overline{\alpha})\alpha}{\overline{\alpha}+(1-\overline{\alpha})\alpha_0}
\left [
\frac{\overline{\alpha}+(1-\overline{\alpha})\alpha_0}{\overline{\alpha}+(1-\overline{\alpha})\alpha}
\,\frac{\alpha-\overline{\alpha}}{\alpha_0-\overline{\alpha}}
\right ]^{1/(2-\overline{\alpha})}\!\!\!
= \frac{\alpha}{\alpha_0} \e^{-\theta},\nn
\end{align}
which (in the model case of constant temperature) even has the analytical solution $\alpha(\theta)$ describing the kinetics between initial condition
$\alpha(0)=\alpha_0$ and final asymptotic equilibrium value $\alpha(\infty)=\overline{\alpha}$.
Let us analyze the relaxation close to equilibrium when $|\alpha-\overline\alpha|\ll1$.
Simple differentiation of the right side of Eq.~(\ref{dot_alpha})
gives approximate equation and its solution
\begin{align}&
\frac{\dm\alpha}{\dm t}
=-\frac{1}{\tau_v}(\alpha-\overline\alpha),\qquad
\frac{1}{\tau_v}\equiv
(2-\overline\alpha)\beta n_\rho,
\label{tau_v}
\\&
(\alpha-\overline\alpha)
=(\alpha_0-\overline\alpha)\e^{-t/\tau_v}.\nn
\end{align}
The subscript emphasizes that this relaxation time $\tau_v$ is calculated at fixed volume
and mass density $\rho.$ 
We will insert so estimated relaxation time in the 
Mandelstam-Leontovich\cite{Mandelstam:37} 
consideration of the second viscosity.\cite{LL6}
\textcolor{black}{In order to consider non-equilibrium thermodynamics 
we solved the kinetic equation for the dime dependent degree of ionization $\alpha(t)$.
The theory Mandelstam-Leontovich\cite{Mandelstam:37} theory is devoted 
on the influence of chemical reactions on the hydrodynamics and creation
by the chemical reactions second viscosity is convincingly described 
in Section~81 ``Second viscosity'' in the 6th volume of the Landau-Lifshitz
course on theoretical physics.\cite{LL6}
The variables $\xi_n$ and $\tau_n$ can be considered as describing chemical compounds
and relaxation times of different chemical reactions in a reacting mixture; definitely no single variables.
The whole monograph by Rudenko and Soluyan\cite{Rudenko} analyzes related topics.
Confer also with the general description of non-equilibrium thermodynamic
by De~Groot and Masur.\cite{DeGroot}
}

In the present paper for illustration we focus on the high-frequency case
for which $\omega\gg \tilde \nu_\beta=\beta n_\rho$ which gives frequency independent absorption.
For these high frequencies for one wave period ionization-recombination process is negligible
and we can apply perturbative consideration of the entropy production.

Let us recall the general formula for the heat of the chemical reaction
applied to the Saha equation for atomic hydrogen.\cite{LL5}
Substitution of temperature dependent constant of of chemical equilibrium
from Eq.~(\ref{Saha}) gives
\be
\delta Q_p=-T^2 n_e\frac{\partial \ln K_p(T)}{\partial T}
=(I+c_pT)n_e, \quad c_p \equiv \frac52.
\ee
This formula has very simple sense. For every electron 
we have first to detach from the atom spending ionization potential $I$,
then $c_pT$ is the enthalpy per free electron and electron density is $n_e$.
In this case $\delta Q_p=\Delta H$ is the change of the enthalpy $H$ per unit volume 
if the ionization is performed at constant pressure $p$.

We also recall the expressions for the volume densities 
of the internal energy
\begin{align}
\mathcal{E}(T,\rho,\alpha) &= c_v (n_e + n_p + n_0 ) T + n_e I\nn\\
&=c_v(1+\alpha)Tn_\rho+\alpha\,n_\rho I.
\label{energy}
\\&
\left(
\frac{\partial\mathcal{E}}{\partial\alpha}
\right)_{T,\rho}
=(c_vT+I) n_\rho,\qquad c_v \equiv \frac32,
\end{align}
and entropy 
\begin{align}
\mathcal{S}(T,\rho,\alpha) &
=\mathcal{S}_e+\mathcal{S}_p+\mathcal{S}_0,
\label{formulae_for_entropy}\\
\mathcal{S}_e&
=n_e \left ( \ln \left [ \frac{g_en_{_q}}{n_e}  \right ] 
+ c_v +1\right ),\\
\mathcal{S}_p&
=n_p \left ( \ln \left [ \frac{g_pn_{_Q}}{n_p}  \right ] 
+ c_v +1\right ), \\
\mathcal{S}_0&
=n_0 \left ( \ln \left [ \frac{g_0n_{_Q}}{n_0}  \right ] 
+ c_v +1\right ),\\&
\left(
\frac{\partial\mathcal{S}}{\partial\alpha}
\right)_{T,\rho}
=\left(\ln\left[\frac{n_qn_0}{n_en_p}
\right]
+c_v
\right)n_\rho,
\end{align}
\noindent
where $g_0=4$, $g_e=g_p=2$ 
are the statistical weights of the particles.
\textcolor{black}{Here we wish to emphasize that this formula for the energy
Eq.~(\ref{energy}) already contains the energy of the chemical reaction, i.e.
the ionization potential $I$. 
For every electron in the initial Hamiltonian we have energy 
$\varepsilon_{\mathbf{p},e}=p^2/2m+I.$
That is why if we analyze the time derivatives of the energy 
we should not include $\sum_i\mu_i \dot N$ terms, for example,
the chemical energy has already been taken into account.
}

In the next section we derive their kinetics.
Now we can clarify the problem which we are going to solve.
We suppose that the chemical reaction of ionization-recombination is relatively 
slow process and in every moment we have a cocktail of
ideal gases of electrons, protons and neutral atoms with Maxwell velocity distribution
with one and the same temperature $T$.
We suppose that plasma has low density and the correlation energy 
is negligible.
The mass density $\rho$ is created by protons and neutral atoms 
as the electrons have negligible contribution.
We have simple and well-known thermodynamic expressions for the entropy
$\mathcal{S}(T,\rho,\alpha)$,
and energy 
$\mathcal{E}(T,\rho,\alpha)$
per unit mass, and the pressure
$p(T,\rho,\alpha)$
is created by all particles.
The Saha equation Eq.~(\ref{Saha_2}) 
for the equilibrium degree of ionization $\bar\alpha$
is consequence of the condition 
of minimal free energy per unit volume
$\mathcal{F}=\mathcal{E}-T\mathcal{S}$
, i.e. zero derivative of the free energy at $\alpha=\bar\alpha$
\begin{align}&
\left(\frac{\partial\mathcal F}{\partial \alpha}\right)_{T,\rho}
=\left(I-T\ln\left[\frac{n_qn_0}{n_en_p}
\right]
\right)n_\rho =0,\\&
\left(\frac{\partial^2\mathcal F}{\partial \alpha^2}\right)_{T,\rho}
=\frac{2-\alpha}{(1-\alpha)\alpha}Tn_\rho>0.
\end{align}
\indent
The time dependent degree of ionization $\alpha(t)$ however is an
additional variable which determines the state of the hydrogen plasma
and for its determination it is necessary to solve kinetic equation whose parameters 
are determined by the cross-sections of elementary processes in the plasma.
We wish to emphasize that in this situation 
($0<\alpha<1$)
there is no connection between the energy and entropy, i.e. the function $\mathcal{E}(\mathcal{S})$ does not exist. \\ \indent
We suppose that a sound wave with evanescent amplitude
propagates through the plasma and variations of the degree of ionization
$\alpha(t)-\bar{\alpha}$ are relatively small, but in the next section we start with consideration of the general case.

%%%%%%%%%%%%%%%%%%%%%%%%%%%%%%%%%%%%%
\section{Kinetic equation for the entropy and H-theorem for chemical reactions}
   \label{Kinetic equation for the entropy}
Radiation processes are essential only for very low densities.
As was pointed out by Pitaevskii,\cite{LL10} 
for dense enough but still cool $T\ll I$ plasma,   
the main role can come to recombination 
with participation of a second electron 
as a third body.
In this case the chemical equilibrium relation
Eq.~(\ref{balance}) and Saha ionization equation 
Eq.~(\ref{Saha_2}) give the rate of the two electron
recombination
\be
\gamma=\frac{\beta}{n_{_\mathrm{S}}}
=\frac{C_\beta \beta_\mathrm{B}}{n_q(T)}
=C_\beta \beta_\mathrm{B}
 \left ( \frac{2 \pi \hbar^2}{mT}\right )^{\!\! 3/2},
\label{2e_recombination}
\ee
which together with the ionization rate
Eq.~(\ref{ionization_rate}) is one of the results of the present
paper.
In this section we use the approximate kinetic equation
\be
\dot n_e \approx \beta n_0  n_e  - \gamma n_p n_e^2.
\label{dot_electron_concentration}
\ee
\textcolor{black}{
As a lateral comment we wish to add that  
long time ago Burgess and Seaton\cite{Burgess} concluded  that
recombination rates by di-electronic processes for coronal ions
are much larger than hitherto bees supposed.
}
Let in the beginning consider spatial homogeneous
plasma with constant mass $n_\rho=\mathrm{const}$ 
and energy density 
$\mathcal{E}=\mathrm{const}$, 
which is not in chemical equilibrium.
Differentiation of Eq.~(\ref{energy}) gives for the time derivative of the temperature
\be
c_v n \dot T = - (I+c_v T) \dot{n}_e, \qquad \dot T \equiv \dm T/\dm t.
\label{time_derivative_of temperature}
\ee
This time derivative we have to substitute 
in the time derivative of the entropy
\be
\dot{\mathcal{S}}
=\frac{\partial \mathcal{S}}{\partial T}\,\dot T
+\frac{\partial \mathcal{S}_e}{\partial n_e}\,\dot n_e
+\frac{\partial \mathcal{S}_p}{\partial n_p}\,\dot{n}_p
+\frac{\partial \mathcal{S}_0}{\partial n_0}\,\dot{n}_0.
\ee
\textcolor{black}{
An elementary substitution here of the derivatives of the formulae 
for the entropy Eq.~(\ref{formulae_for_entropy}),
time derivative of the temperature Eq.~(\ref{time_derivative_of temperature})
and time derivative of the electron concentration 
expressed by the kinetic equation Eq.~(\ref{dot_electron_concentration})
}
after some algebra gives the time derivative of the entropy
\begin{align}\label{Eta_chemistry}
\dot{\mathcal{S}}=&
\left(
\beta n_0 n_e-\gamma n_p n_e^2
\right)
\ln\left[\dfrac{\beta n_0n_e}{\gamma n_pn_e^2}\right]
\\=&(\beta n_e n_0)
\ln \!\! \left [
 \frac{n_e n_p}{n_0n_{_\mathbf{S}}}\right ] \!
\left(\frac{n_e n_p}{n_0n_{_\mathbf{S}}}-1\right)
\nn\\
=& (\mathcal{X}-\mathcal{Y})
\ln\frac{\mathcal{X}}{\mathcal{Y}}\ge 0,\nn\\
\mathcal{X}\equiv&\,\dot n_e^{(+)}=\beta n_0n_e>0,\qquad
\mathcal{Y}\equiv\dot n_e^{(-)}=\gamma n_pn_e^2>0;\nn
\end{align}
\textcolor{black}
{we have doubled the notations in order to alleviate the understanding of the entropy increase.}

Introducing
a dimensionless variable 
which describes the deviation from the chemical equilibrium
\begin{align}&
\chi \equiv\frac{n_e n_p}{n_0n_{_\mathbf{S}}}-1
=\frac{n_e n_p}{n_0}\frac{\bar n_0}{\bar n_e \bar n_p}-1
\label{chi}
\end{align}
and positive income rate in the kinetic equation
\be
\dot n_e^{(+)}\approx\beta n_0 n_e>0
\ee
the entropy production reads as
\be
\dot{\mathcal{S}}=\dot n_e^{(+)}\,\eta(\chi),\quad
\eta(\chi)\equiv\chi \ln(1+\chi)>0,
\ee
which together with Eq.~(\ref{Eta_chemistry}) is actually the general Boltzmann Eta-theorem 
applied to chemical reactions in gaseous phase.
We have quite general expression for the entropy production
per unit volume $\dot{\mathcal{S}}$ in fluids:
difference of the income $\mathcal{X}$ 
and outcome $\mathcal{Y}$ terms times 
the logarithm of their ratio.
Our hypotheses is that this result can be extended even for chemical reactions 
in liquid solutions, and for the general case we have to perform the summation 
$\sum_r$
over all reactions (denoted by index "$r$") in the fluid
\be
\frac{\mathrm{d}\mathrm{H}}{\mathrm{d}t}
=\sum_r (\mathcal{X}_r-\mathcal{Y}_r)\ln\frac{\mathcal{X}_r}{\mathcal{Y}_r}.
\ee
Coining international word for the measure of chaos
Klausius made a compound of Greek preposition $\eta\nu$ akin to English
preposition ``in'' and Greek word $\tau\rho\mathrm{o}\pi o \varsigma$
(say aspiration) revealing inherent aim of every system for disorder.
English readers of Klausius and Boltzmann articles pronounce capital Greek 
$\eta$ (Eta) as Latin``H''. Now nobody remember this old story.
In our current problem the increase of entropy is
due to ionization-recombination processes in hydrogen plasma.

For many practical problems it is important to know the 
entropy production of a system close to equilibrium.
For small deviations of the concentrations from the Saha 
equilibrium equation
\be
\eta(\chi)= \chi \ln(1+\chi)\approx \chi^2, 
\quad\mbox{for}\quad |\chi|\ll1
\ee
we can take constant income rate in equilibrium
\be
\nu \equiv \beta \bar n_0 \bar n_e\equiv \bar{\mathcal{X}}=\mathcal{\bar{Y}}
\ee
and the volume density of entropy production takes the form
\be
\dot{\mathcal{S}}\approx\nu\chi^2,
\quad 
\mathcal Q=T\dot{\mathcal{S}}
=Q_\mathrm{\iota}\chi^2,\quad
Q_\mathrm{\iota}\equiv T\nu.
\label{dot_S}
\ee
This small increase of the entropy at almost constant temperature is related to 
\textcolor{black}{the heat production per unit volume and time,} i.e. the power density of the irreversible 
processes $\mathcal Q$ parameterized by
a constant $Q_\mathrm{\iota}$ describing the rate of the chemical reactions per unit volume.
\textcolor{black}{For systems close to equilibrium, with constant temperature, 
the absorbed heat energy is given by the product of temperature and increase of entropy.}

In the next section we apply average volume density of dissipated power 
\be
\overline{\mathcal{Q}}
=Q_\mathrm{\iota}\left<\chi^2\right>,
\label{power}
\ee
where brackets denote time averaging,
for the problem of damping of sound waves
in partially ionized hydrogen plasma.
This averaging of energy production on the period of oscillations
is a good approximation only for high enough frequencies
\be\omega\gg \tilde \nu_\beta\equiv(1-\bar\alpha)\bar\alpha\,\beta n_\rho
\label{preliminary_criterion}\ee
for which the change of the ionization degree is small
for the period of oscillations.
Later we will precise this result later.
It seems strange that this general formula for the entropy production
$\dot{\mathcal{S}}=\nu \,\chi \ln(1+\chi)$
can be found perhaps for the first time in the present work.
For small concentrations and densities the ratio $\nu/\prod c$
gives the rate of the chemical reactions poorly accessible by other methods.

%%%%%%%%%%%%%%%%%%%%%%%
\section{High frequency sound absorption}
   \label{Sound absorption}
In this section we derive the wave damping rate applying standard method 
for calculation of damping rate for high quality oscillations.
The first step is to calculate the energy loss by dissipation per unit period\cite{LL1}
and in such a way we obtain the obvious relation
\be
-\dot{\overline{\mathcal{E}}}=T\dot{\overline{S}}= \overline Q
\ee
describing that the averaged loss of the mechanical energy per unit time
$\dot{\overline{\mathcal{E}}}$ is equal to the averaged heating power $\overline Q$, which for a fluid with temperature $T$ is proportional to the time averaged entropy production $\dot{\overline{S}}$.
The averaged chemical compound is not changed $\overline\alpha(t)$.
This method is widely used for calculation of wave damping\cite{LL6} 
for gravitational waves in deep water, narrow channels, sound waves and influence of chemical reactions on the second viscosity, etc.\cite{LL6}
Wave damping is created by the irreversible heating of the fluid.
In magnetohydrodynamics alongside the mechanical energy, it is necessary to add the energy of the magnetic field,\cite{LL8} but the general scheme works without any modification.
Let us consider a plane sound wave 
with frequency $\omega$ and 
$x$-axis chosen along the wave-vector $\mathbf{k}$
and the direction of the longitudinal
oscillation of the velocity\cite{LL6}
\begin{align}
v_x(x,t)=v_0\cos(kx-\omega t),\quad
\omega= c_s k,\quad
v_0\ll c_s
\end{align}
with amplitude $v_0$ much smaller than the sound speed 
$c_s$.
The sound wave has the averaged volume density 
of mechanical energy 
\begin{align}
\overline{\mathcal{E}}
=2\left<\frac12 \rho v^2 \right>=\frac{\rho v_0^2}{2},
\label{E_density}
\end{align}
twice time the averaged density of the kinetic 
energy density $\rho v^2/2$ 
and $\left<\cos^2\right>=1/2.$
The calculated in the former section
dissipation power is just the damping power of the sound wave
\be
\overline{\mathcal{Q}}=-\dot{\overline{\mathcal{E}}}.
\label{power_density}
\ee
According H-theorem Eq.~(\ref{Eta_chemistry}) the time derivative of the entropy 
$\dot{\overline{\mathcal{S}}}$ is positive
and for small wave amplitude it is quadratic Eq.~(\ref{dot_S}) 
with respect of the wave amplitude.
Supposing that amplitude of the velocity 
$v_0(t)\propto\e^{-\gamma_t t}$ is exponentially decaying 
with small rate $\gamma_t\ll\omega$.
For the energy density the rate is doubled
$\overline{\mathcal{E}}\propto\e^{-2\gamma_t t}$.
The damping rate is given by the logarithmic derivative
\be
2\gamma_t
=-\frac{\dot{\overline{\mathcal{E}}}}{\overline{\mathcal{E}}}
=\frac{\overline{\mathcal{Q}}}{\langle\mathcal{E}\rangle}.
\ee
For sound waves which have negligible dispersion
$c_s=\omega/k$
it is more convenient to follow 
a wave packet propagating along the $x$-axis
with coordinate of the packet (in quasi-classical  approximation) 
$x=c_st$.
This means that we consider spatial dependence of the 
amplitude of the velocity 
$v_0\propto\e^{-\gamma_\mathrm{ion} x}$
and for the space damping rate we have
\be
\gamma_\mathrm{ion}
=\frac{\overline{\mathcal{Q}}}{2c_s\langle\mathcal{E}\rangle};
\label{gamma_ion}
\ee
\textcolor{black}
{
the argument of the exponential functions are dimensionless 
and this gives the conditionality of the damping rates $\gamma_\mathrm{ion}$
and $\gamma_t.$
}
In order to calculate this ionization induced damping
we have to calculate the numerator 
$\overline{\mathcal{Q}}$
considering plane sound wave.

Oscillations of the velocity $v(x,t)$
create oscillations of the 
mass density $\rho=M n_\rho$, 
temperature $T$, electron $n_e$, proton $n_p$ and
atom $n_0$ densities, and the density of Saha 
$n_{_\mathrm{S}}$ given by definition Eq.~(\ref{Saha}):
\begin{align}
&
\rho=\rho_0+\rho^\prime,\quad
n_\rho=n_{\rho,0}+n_\rho^\prime,\quad
n_0=\bar n_0+n_0^\prime,
\\
&
n_e=\bar n_e+n_e^\prime,\quad
n_p=\bar n_p+n_p^\prime,\quad
T=T_0+T^\prime,\quad\\
&
n_{_\mathrm{S}}(T)=n_{_\mathrm{S}}(T_0)
+\frac{\dm n_{_\mathrm{S}}}{\dm T}T^\prime,\\
&
\frac{\dm n_{_\mathrm{S}}}
{n_{_\mathrm{S}}\dm T}
=\frac{I+c_vT}{T^2}= \frac{\iota+c_v}{T}\approx \frac{\iota}{T},
\quad\mbox{for}\quad \iota=I/T\gg1,\nn
\end{align}
where the 0 subscript is omitted from now on in the linearized approximation for all oscillating variables.\cite{LL8}

The mass conservation equation 
$\partial_t \rho=-\mathrm{div}(\rho \mathbf{v})$
for small amplitude waves gives
\be
\frac{\rho^\prime}{\rho}=\frac{v_0}{c_s}\sin(kx-\omega t).
\ee
For adiabatic in initial approximation compression
the chemical compound is not changed and
all particles have proportional oscillations of the density.
\be
\frac{n_e^\prime}{\bar n_e}
=\frac{n_p^\prime}{\bar n_p}
=\frac{n_0^\prime}{\bar n_0}
=\frac{n_\rho^\prime}{\bar n_\rho}
=\frac{\rho^\prime}{\rho}
=\frac{v_0}{c_s}\sin(kx-\omega t).
\ee

Here we emphasize that for high frequencies $\omega\gg \tilde \nu_\beta$
the influence of the chemical reaction is small and in a good approximation
hydrogen plasma can be considered for monoatomic gas with constant
heat capacities and atomic polytropic index $\gamma_a=5/3.$
For adiabatic $S=\mathrm{const}$ compression 
of an ideal gas with constant heat capacity per particle
$c_v=3/2$
we have the well known law\cite{LL5} 
$TV^{1/c_v}=\mathrm{const}$
whence
\be
\frac{T^\prime}{T}=\frac1{c_v}\frac{\rho^\prime}{\rho},
\ee
This oscillation of the temperature describes the oscillations
of the temperature density determining the ionization rate
\be
\frac{n_{_\mathrm{S}}^\prime}
{n_{_\mathrm{S}}}
\approx \frac{I}{T}\frac{T^\prime}{T}
=\frac{\iota}{c_v}\frac{\rho^\prime}{\rho}.
\ee

Now we have all ingredients to calculate
the variable $\chi$ from Eq.~(\ref{chi}) describing 
the deviation from equilibrium
\be
\chi \approx
\frac{n_e^\prime}{\bar n_e}
+\frac{n_p^\prime}{\bar n_p}
-\frac{n_0^\prime}{\bar n_0}
-\frac{n_{_\mathrm{S}}^\prime}
{n_{_\mathrm{S}}}
\approx
\frac{\iota}{c_v}\frac{n_\rho^\prime}{\bar n_\rho}
=\frac{\iota v_0}{c_v c_s} \sin(kx-\omega t). \nn
\ee
For low temperatures $T\ll I$
the dimensionless ratio $\iota=I/T\gg1$
and only variations of the Saha density $n_{_\mathrm{S}}$ 
are essential.
Averaging $\left<\sin^2\right>=1/2$ gives
\be
\left<\chi^2\right>
=\frac12\left(\frac1{c_v}\frac{Iv_0}{Tc_s}\right)^2\ll1.
\ee
Now we can calculate the power according
Eq.~(\ref{power})
\be
\left<\mathcal{Q}\right>
=\frac12\left(\frac{\iota \, v_0}{c_vc_s}\right)^2Q_\mathrm{\iota}.
\ee
Substitution of this power density
together with energy density
Eq.~(\ref{E_density}) 
in the definition for the extinction 
Eq.~(\ref{gamma_ion}) 
finally gives
\begin{align}
\gamma_\mathrm{ion}
& =\frac{Q_\mathrm{\iota}\iota^2}{2c_v^2\rho c_s^3}
=C_\beta\beta_\mathrm{B}\,\frac{v_{Tp}^2}{2c_v^2c_s^3}
\,(1-\alpha)\alpha \,n_\rho\, \frac{I^2}{T^2}\,\e^{-I/T},
\label{ion_and_usual_hydro}\\
\gamma_\mathrm{tot}&\!=\!\frac{\omega^2}{2\rho\, c_s^3}\left[
\left(\frac43\eta+\zeta^\prime\right)
+\left(\frac1{\mathcal{C}_v}\!-\!\frac1{\mathcal{C}_p}\right)\!\varkappa
\right],
\label{total_extinction}\\
&\gamma_\mathrm{tot}=\gamma_\eta+\gamma_\zeta+\gamma_\varkappa,\quad
\gamma_\eta\equiv\frac{\omega^2}{2\rho\, c_s^3}\,\frac43\eta,\nn\\
&\gamma_\zeta(\omega)\equiv\frac{\omega^2}{2\rho\, c_s^3}\,\zeta^{\prime}(\omega),\quad
\gamma_\varkappa\equiv\frac{\omega^2}{2\rho\, c_s^3}\,
\left(\frac1{\mathcal{C}_v}\!-\!\frac1{\mathcal{C}_p}\right)\!\varkappa,\nn\\
& \gamma_\mathrm{ion}=\gamma_\zeta(\omega\gg\tilde\nu_\beta),\nn
\end{align}
where we have introduced the thermal velocity of protons $v_{Tp} \equiv \sqrt{T/M}$.
For comparison we have also introduced the well-known result for the damping of 
sound waves\cite{LL6} where
$\mathcal{C}_v$ and $\mathcal{C}_p$
are heat capacities of the partially ionized hydrogen plasma per unit mass 
recently calculated:\cite{gamma_hydrogen}
\begin{align}&
\tilde{c}_p\equiv
\frac{\rho\,\mathcal{C}_p}{n_\mathrm{tot}}
=c_p+(c_p+\iota)^2\varphi,\label{c_p}\\&
\tilde{c}_v\equiv\frac{\rho\,\mathcal{C}_v}{n_\mathrm{tot}}
=c_v+(c_v+\iota)^2\varphi/(1+\varphi),\label{c_v}\\&
\tilde\gamma=\dfrac{\mathcal{C}_p}{\mathcal{C}_v}
=\frac{c_p+(c_p+\iota)^2 \varphi}
{c_v+(c_v+\iota)^2 \varphi/(1+\varphi)},
\label{gamma}\\&
\varphi\equiv\frac12(1-\alpha)\alpha, \quad 
\frac{\rho}{n_\mathrm{tot}}\equiv\left<M\right>\approx\frac{M}{1+\alpha},
\label{averaged_mass}
\end{align}
where $\left<M\right>$ is the averaged mass of the cocktail,
and $\tilde{c}_v$ and $\tilde{c}_p$ are temperature and ionization dependent 
heat capacities per particle; the temperature is in energy units.
for pure hydrogen plasma,
$\varkappa$ is the heat conductivity,
$\eta$ and $\zeta^\prime$ are respectively the first and second viscosity coefficients.
The subscript ``ion'' means ionization-recombination term of the extinction,
the imaginary part of the wave-vector.

Often if 
\be
\alpha\left(\frac{e^2}{T}\right)^{\!2}\! \mathcal L\gg a_\mathrm{B}^2,
\quad \mathcal L= \ln\!\left(\frac{a_\mathrm{_D} T}{e^2}\right),\quad 
\frac1{a_\mathrm{_D}^2}=\frac{4\pi e^2n_\mathrm{tot}}{T}
\ee
for the partially ionized plasma one can use with an acceptable approximation the
the estimations for completely ionized hydrogen plasma\cite{LL5,LL10}
\be
\eta\approx 0.4\, M^{1/2}T^{5/2}/e^4\mathcal{L},\quad
\varkappa\approx 0.9\, T^{5/2}/e^4\mathcal{L}m^{1/2}
\ee
which are density independent.
For weakly or almost completely completely ionized hydrogen plasma
when $\tilde c_p\approx c_p$
the Prandtl\cite{LL6} number is temperature independent
\begin{align}
\mathrm{P}\equiv\dfrac{\eta/\rho}{\varkappa/\rho \mathcal{C}_p}
\sim \sqrt{\frac{m}{M}}\ll 1.
\end{align}

The sound velocity is determined by the ratio of the two heat
capacities $\gamma_a$ and the averaged mass of the gas cocktail $\left<M\right>$
\begin{align}&
c_s=\sqrt{\gamma_a T/\left<M\right>},
\qquad \gamma_a\equiv c_p/c_v,
\\&
\left<M\right>
=\frac{m n_e+M n_p+Mn_0}{n_e+n_p+n_0}.
\end{align}
Expressing the concentrations by the ionization degree from \Eqref{pressure} and \Eqref{n_alpha}
we obtain Eq.~(\ref{averaged_mass})
\be
\left<M\right>=M/(1+\alpha), \qquad m\ll M.
\ee
Additionally substituting $c_v=\frac32$, $\gamma_a=\frac53$ and $v_{Tp}$
the sound speed can be expressed as
\be
c_s=\sqrt{\gamma_a}\,v_{Tp}\,\sqrt{1+\alpha}.
\ee
The total extinction 
$\gamma_\mathrm{tot}$ is just the sum
the hydrodynamic one proportional to the shear viscosity $\eta$,
the bulk or compression viscosity $\zeta$
created by ionization-recombination chemical reactions
and the contribution of the heat conductivity $\varkappa$.
The formulae are valid 
for sound wave-vectors 
smaller than total extinction
\be
k=\omega/c_s\gg \gamma_\mathrm{tot}.
\ee
For $\omega\ll \omega_c= c_s \gamma_\mathrm{tot}$
experimental study could give the result
on $\gamma_\mathrm{tot}$ versus $\omega^2$ plot. 
The linear regression will distinguish
the influence of the chemical reactions
$\gamma_\mathrm{ion}$ 
and the hydrodynamic coefficient $a_\eta$.
The extinction is actually the imaginary
part of the wave-vector
\be
k=k_1+\mathrm{i}k_2,\quad
\e^{\mathrm{i}kx}
=\e^{\mathrm{i}k_1x}
\e^{-k_2x},\quad
k_2=\gamma_\mathrm{tot}
\ee
Again in the quasi-classical  approximation,
for the high frequency sound waves propagating along the $x$-direction
the extinction represents 
the spatial dependence of the energy flux $q_s(h)$ of sound waves
\begin{align}
q_s(h)=q_s(0)\exp
\left(-\int_0^h 2\gamma_\mathrm{tot}(x)\,\dm x\right).
\label{energy flux}
\end{align}\\

%%%%%%%%%%%%%%%%%%%%%%%%%%%%%%%%%
\section{Mandelstam-Leontovich theory for sound propagation}
    \label{M-L}

In this section following Landau and Lifshitz\cite{LL6} 
we apply the the Mandelstam-Leontovich\cite{Mandelstam:37}
theory for the influence of slow chemical reactions on the sound propagation.
The simplicity of ionization-recombination processes in hydrogen allows 
analytical treatment with a satisfactory precision.

In order to derive the second viscosity coefficient of the hydrogen plasma
we use the Mandelstam-Leontovich\cite{Mandelstam:37} 
results as they are presented by Landau-Lifshitz.\cite{LL6}
The main details are the derivatives of the pressure with respect to the mass density.
There are two limit cases: 1) fast adiabatic compression of mono-atomic gases 
at constant degree of ionization $\alpha$ and 2)
extremely slow compression at which in every moment the ionization degree 
follows the equilibrium Saha equation corresponding to the density
\begin{align}
c_\infty^2\equiv c_s^2
=\left(\frac{\partial p}{\partial \rho}\right)_{\!\!\alpha},
\quad
 c_0^2
\equiv\left(\frac{\partial p}{\partial \rho}\right)
_{\!\!\mathrm{eq}},
\quad
c_0<c_\infty.
\end{align}
Here we re-denote $c_\infty\equiv c_s$ and from now on $\alpha \equiv \overline \alpha$,
i.e. we work with equilibrium ionization degree.

%%%%%%%%%%%%%%%%%%%%%%%%%%%%%%%%%
\subsection{Calculation of equilibrium compressibility $c_0^2$}
In order to calculate the equilibrium compressibility 
we will trace the change of the energy of a ``liquid particle''\cite{LL6}
with volume $V$ according Eq.~(\ref{energy}) and Eq.~(\ref{pressure})
we have
\be
E=(c_vp+I\alpha n_\rho)V.
\ee
When a sound wave with evanescent frequency $\omega\rightarrow 0$ 
propagates through the plasma the compression is without 
external heat and the energy conservation gives 
$\mathrm{d}E=-p\,\mathrm{d}V.$
Now we can calculate the change of the volume density of the energy
$\mathrm{d}E/V$.
Additionally mass conservation $\rho V =\mathrm{const}$ gives
\be
\frac{\mathrm{d}V}{V}
=-\frac{\mathrm{d}\rho}{\rho}
=-\frac{\mathrm{d}n_\rho}{n_\rho}
\ee
and the equation for the $\mathrm{d}E/V$ gives
\be
I n_\rho \frac{\mathrm{d}\alpha}{p}
=\left(c_p\frac{\mathrm{d}n_\rho}{n_\rho}
-c_v\frac{\mathrm{d}p}{p}\right).
\ee
From the equation for the pressure Eq.~(\ref{pressure}) we can express 
$n_\rho=p/(1+\alpha)T$ which together with the definition
for $\iota$ Eq.~(\ref{iota}) substituted in the equation above gives
\be
\mathrm{d}\alpha
=\frac{1+\alpha}{\iota}
\left(c_p\frac{\mathrm{d}n_\rho}{n_\rho}
-c_v\frac{\mathrm{d}p}{p}\right).
\label{d_alpha}
\ee
Now we can differentiate the equation for the pressure
Eq.~(\ref{pressure}) and $\mathrm{d}[p=(1+\alpha)n_\rho T]$
gives
\be
\frac{\mathrm{d}p}{p}
=\frac{\mathrm{d}T}{T}
+\frac{\mathrm{d}n_\rho}{n_\rho}
+\frac{\mathrm{d}\alpha}{1+\alpha}.
\ee
The substitution here $\mathrm{d}\alpha$ from Eq.~(\ref{d_alpha}) gives
\be
\frac{\mathrm{d}T}{T}=\left(1+\frac{c_v}{\iota}\right)\frac{\mathrm{d}p}{p}
-\left(1+\frac{c_p}{\iota}\right)\frac{\mathrm{d}n_\rho}{n_\rho}.
\label{dT/T}
\ee
The last detail from this mosaic is the differentiation of the Saha equation
in the form Eq.~(\ref{p_alpha_n}) which gives
\be
\frac{p}{p_{_\mathrm{S}}}\frac{\mathrm{d}p}{p}
-\frac{p}{p_{_\mathrm{S}}}\frac{\mathrm{d}p_{_\mathrm{S}}}{p_{_\mathrm{S}}}
=-\frac{2}{\alpha^3}\,\mathrm{d}\alpha.
\label{d_Saha}
\ee
The elementary differentiation of Eq.~(\ref{Saha}) gives
\begin{align}
\frac{\mathrm{d}p_{_\mathrm{S}}}{p_{_\mathrm{S}}}
=\left(c_p+\iota \right)\frac{\mathrm{d}T}{T},
\end{align}
and we substitute it in Eq.~(\ref{d_Saha}).
Then this equation Eq.~(\ref{d_Saha}) we substitute $\mathrm{d}T/T$
from Eq.~(\ref{dT/T}) and $\mathrm{d}\alpha$ from Eq.~(\ref{d_alpha}),
and also $\mathrm{d}n_\rho/n_\rho=\mathrm{d}\rho/\rho.$
The so modified Eq.~(\ref{d_Saha}) 
we multiply with $\iota\alpha^3/(1+\alpha)$ and 
select terms multiplied by $\mathrm{d}p$ and $\mathrm{d}\rho$.
After some algebra we finally derive
\begin{align}
c_0^2&\equiv\left.\frac{\mathrm{d}p}{\mathrm{d}\rho}\right|_\mathrm{eq}
=\frac{2c_p+(1-\alpha)\alpha\,(c_p+\iota)^2}
{2c_v+(1-\alpha)\alpha\,[(c_v+\iota)^2+c_v]}
\frac{p}{\rho}\\
&=\frac{1}{1+\varphi}\frac{\tilde c_p}{\tilde c_v}\frac{p}{\rho}.
\label{c_0^2}
\end{align}
Here index "eq" denotes that compression is so slow that in every moment
the ionization degree $\alpha(t)$ follows the equilibrium ionization given by the Saha equation for slowly changing mass density .
For comparison we rewrite the adiabatic compressibility corresponding 
for such high frequencies (formally $\omega\rightarrow\infty$) for which 
the chemical compound is constant (and informally we can call this process ``iso-chema''),
\be
c_\infty^2\equiv c_s^2
=\left.\frac{\mathrm{d}p}{\mathrm{d}\rho}\right|_\mathrm{s,\alpha}
=\gamma_a \frac{p}{\rho}.
\label{c_infty^2}
\ee
The constants $c_v$, $c_p$ and $\gamma_a$ are just notations from the theory
of mono-atomic gases.

Now we can analyze this final result in different physical conditions:
1)
For $\alpha(1-\alpha)=0$ (more precisely for $\alpha(1-\alpha)\iota^2\ll1$) 
when we have atomic hydrogen or ``mono-atomic''
cocktail from electrons and protons, we have 
$\left. c_0 \right\vert_{\bar\alpha=0} =\left. c_0 \right\vert_{\bar\alpha=1}
\approx c_\infty.$
2)
In the opposite case of $\alpha(1-\alpha)\iota^2\gg1$
we have 
\be
\left. c_0^2 \right\vert_{\iota=\infty} \approx \frac{p}{\rho}.
\ee
It is remarkable that for partially ionized plasma, say $\alpha=1/2$
in the lower solar corona when $T\ll I$
after 300 years we recover the 
the original Newton\cite{Newton} result 
$c_0\approx\sqrt{p/\rho}$, for $\omega\ll 1/\tau.$
In both the cases indexed with ``$\infty$'' and ``eq'' there is no external heating
applied to the gases and in this case we consider adiabatic processes.

Following Rudenko and Soluyan\cite{Rudenko} it is instructive 
to introduce the dimensionless parameter
\begin{align}
\mathcal{M}(\iota,\alpha)&\equiv\frac{c_\infty^2-c_0^2}{c_0^2}
= \frac{\iota^2(1-\alpha)\alpha}
{2c_vc_p+c_v[\iota+c_p]^2(1-\alpha)\alpha}
\nn \\ 
&=\frac{\iota^2\varphi}{c_v\tilde c_p}, \quad
c_v=\frac1{\gamma_a-1}, \quad
c_p=\frac{\gamma_a}{\gamma_a-1},
\end{align}
where we use the notation $\mathcal{M}$ for their parameter $m$ 
defined in page 85 of 
Ref.~\onlinecite{Rudenko}.
For liquids $\mathcal{M}\ll1$ but for hydrogen plasma
our exact result for $\iota^2(1-\alpha)\alpha\gg1$
gives
\be
\mathcal{M}(\iota=\infty)=\gamma_a-1=\frac1{c_v}=\frac23\approx0.666.
\ee
The non-physical case of negative $\iota$ and detachment energy $I$
would correspond to the influence of meta-stable states,
imagine the influence of Wannier ridge resonances 
on the absorption of sound waves in plasmas.

%%%%%%%%%%%%%%%%%
\subsection{Sound propagation}

Supposing that $\omega$ is real for a complex wave-vector we
have the Landau-Lifshitz\cite{LL6} expression
\begin{align}
k=k_1+\mathrm{i}k_2
=\omega\sqrt{\frac{1-\mathrm{i}\omega\tau}
{c_0^2-c_\infty^2\mathrm{i}\omega\tau}},\quad
k_1(\omega)=\Re(k).
\end{align}
Wave propagation requires the imaginary part of the wave-vector to be small  
\be
k_2(\omega)\equiv\gamma(\omega)=\Im(k)\ll k_1=\omega/c_\omega.
\ee
This equation is the definition of 
the sound speed as ration between the frequency and real part of the wave vector $c_\omega\equiv\omega/\Re(k).$
The Mandelstam-Leontovich dispersion has two limit cases\cite{LL6}
\be
k\approx
\frac{\omega}{c_0}
+\mathrm{i}\,\frac{\omega^2\tau}{2c_0^3}
(c_\infty^2-c_0^2) 
=\frac{\omega}{c_0}\left(1+ \frac{\mathrm{i} \omega \tau}{2}\mathcal{M}\right)
\ee
for $\omega\tau\ll1$ and
\begin{align}&
k\approx
\frac{\omega}{c_\infty}
+\mathrm{i}\,\frac{c_\infty^2-c_0^2}
{2\tau c_\infty^3} =
\frac{\omega}{c_\infty}\left(1+ \frac{\mathrm{i}\mathcal{N}}{2\omega \tau}\right),
\quad\omega\tau\gg1,\\
&
\mathcal{N}\equiv\frac{c_\infty^2-c_0^2}{c_\infty^2}=\frac1{c_p}
\frac{\iota^2 (1-\alpha)\alpha}
{2c_v+(1-\alpha)\alpha\,[(c_v+\iota)^2+c_v]}\nn \\&
\quad=\frac{\iota^2}{c_p \tilde c_v}\frac{\varphi}{1+\varphi}.\nn
\end{align}
For arbitrary frequencies, introducing frequency dependent second 
viscosity\cite{LL6}
\begin{align}&
\zeta(\omega)=\zeta^\prime+\mathrm{i}\zeta^{\prime\prime}
\equiv\frac{\zeta_0}{1-\mathrm{i}\omega\tau}
=\zeta_0\frac{1+\mathrm{i}\omega}{1+\omega^2\tau^2},\label{frequency_dependence}\\&
\zeta^\prime(\omega)\equiv\Re(\zeta)
=\frac{\zeta_0}{1+\omega^2\tau^2}=\frac{\zeta_0}{1+f^2/f_\iota^2},\\
& f\equiv\frac{\omega}{2\pi},\qquad f_\iota\equiv\frac1{2\pi\tau},\\
&\zeta_0\equiv \zeta(\omega \ll 1/\tau) = (c_\infty^2-c_0^2)\tau\rho,\label{second_viscosity}\\&
\zeta_\infty(\omega)=\zeta^\prime(\omega\gg1/\tau)
= (c_\infty^2-c_0^2)\rho/\tau\omega^2,
\end{align}
one can use the usual hydrodynamic relation between the second viscosity and the related by it extinction
\be
c_\omega=\frac{\omega}{k_1},\qquad
\gamma_\zeta(\omega)=\frac{\omega^2\Re(\zeta)}{2\rho c_\omega^3},
\ee
cf. Ref.~\onlinecite{Zavershinskii} and references therein.
This general formula gives
quadratic dependence of the extinction $\gamma_\zeta(\omega)\propto\omega^2$ 
at low frequencies $\omega \tau \ll1$, 
and the frequency independent extinction 
$\gamma_{\zeta,\infty}=\gamma(\omega\rightarrow\infty)$ 
for high frequencies
$\omega\tau\gg1$
\begin{align}&
\gamma_\zeta(\omega\rightarrow 0)=\frac{\omega^2\zeta_0}{2\rho c_0^3},\\&
\gamma_{\zeta,\infty}=\frac{c_\infty^2-c_0^2}
{2\tau c_\infty^3}=\frac{\zeta_0}
{2\tau^2\rho c_\infty^3}.
\label{gamma_infty}
\end{align}

We have performed perturbative calculation of 
constant chemical part of the extinction $\gamma_\mathrm{ion}$
analyzing entropy production Eq.~(\ref{ion_and_usual_hydro}) at high frequencies
$\omega\gg1/\tau.$
Now we can determine the time constant $\tau$ of the Mandelstam-Leontovich theory
$\tau$ comparing these two results:
first expression in Eq.~(\ref{gamma_infty}) 
and Eq.~(\ref{ion_and_usual_hydro}) 
$\gamma_\mathrm{ion}\approx\gamma_{\zeta,\infty}$.
The theoretical fit gives
\begin{align}
\tau&
=\frac{c_v(1+\alpha)}{2c_v+(1-\alpha)\alpha\,[(c_v+\iota)^2+c_v]}\frac1{\beta n_\rho}\\&
=\frac{1+\alpha}{2(1+\varphi)}\frac{c_v}{\tilde c_v}\frac1{\beta n_\rho}.
\label{tau_ML}
\end{align}
In such a way for first time we have \textit{ab initio}
calculated parameters $\tau$, $c_0$ and $c_\infty$
of the  Mandelstam-Leontovich theory.\cite{Mandelstam:37} 

The time constant $\tau$ is similar 
but differs from $\tau_v$ defined in Eq.~(\ref{tau_v})
defined by ionization degree relaxation at fixed volume and density $\rho$.
This result for the Mandelstam-Leontovich time constant $\tau$
precises the preliminary criterion Eq.~(\ref{preliminary_criterion}).
\textcolor{black}{
The difference between $\tau_v$ Eq.~(\ref{tau_v}) and $\tau$ Eq.~(\ref{tau_ML}) is natural,
the propagation of sound is definitely not an isochoric process.}
Then we can substitute this time constant in Eq.~(\ref{second_viscosity})
and this is the final result for the second viscosity $\zeta_0$ of partially ionized hydrogen plasma and its frequency dependence $\zeta(\omega)$ Eq.~(\ref{frequency_dependence}).
In this result it is necessary to substitute $c_\infty^2$ from Eq.~(\ref{c_infty^2})
and $c_0^2$ from Eq.~(\ref{c_0^2}) to
obtain the dispersion of second viscosity $\zeta(\omega)$.
The complete formula for the second viscosity $\zeta(\omega)$
which is the central result of the present paper
will be represented in the nest section.
It will be nice if the same result can be re-derived directly from the 
kinetics of plasma without the use 
of Mandelstam-Leontotic theory.

The results for sound damping $\gamma(\omega)$ have
fundamental importance for calculation of heating of solar chromosphere
by sound waves.
The spectral density of the heating function $Q_\omega$
is expressed by the extinction $\gamma_\mathrm{tot}$ and the spectral density of the
energy flux of the longitudinal sound waves $q_\omega$
\be
Q_\omega=2\gamma_\mathrm{tot}(\omega) q_\omega,\quad
Q=\!\int\limits_{-\infty}^\infty \! Q_\omega\frac{\mathrm{d}\omega}{2\pi},\quad
q=\!\int\limits_{-\infty}^\infty \! q_\omega\frac{\mathrm{d}\omega}{2\pi},
\ee
where $q$ is the full energy flux of the sound waves and 
$Q$ is the total heating power density by sound absorption.
The calculation of damping rate MHD waves will be a subject of another work.

In general, the chemical wave damping rates $\gamma_\zeta$ 
will be crucial in our understanding of the heating of stellar atmospheres.

%%%%%%%%%%%%%%%%%%%%%%%%%%%%%%%%%%%%
\section{Second viscosity and and the heating of solar chromospere}

Taking $c_0^2$ from Eq.~(\ref{c_0^2}) and $c_\infty^2$ from 
Eq.~(\ref{c_infty^2}) we calculate the difference
\begin{align}
&
c_\infty^2-c_0^2=\frac{p}{\rho}\mathcal{F}(\iota,\alpha),
\quad v^2_{Tp}=\frac{T}{M},\quad 
\frac{p}{\rho}=(1+\alpha)v^2_{Tp}, \nn \\
&\mathcal{F}\equiv\frac1{c_v}
\frac{\iota^2 (1-\alpha)\alpha}
{2c_v+(1-\alpha)\alpha\,[(c_v+\iota)^2+c_v]}\nn\\&
=\gamma_a\frac{c_0^2}{c_\infty^2}\mathcal{M}
=\frac{\varphi}{1+\varphi}\frac{\iota^2}{c_v\tilde c_v}
.
\end{align}
Then substituting this difference and time constant
$\tau$ from Eq.~(\ref{tau_ML}) in Eq.~(\ref{second_viscosity})
we finally arrive at
\begin{align}
&
\zeta_0=\frac{c_v(1+\alpha)^2}{a_v}\,\frac{T}{\beta}\,
\mathcal{F}(\iota,\alpha)
\!=\!\frac{\varphi}{2}\frac{(1+\alpha)^2}{(1+\varphi)^2}\frac{\iota^2}{\tilde c_v^2}\frac{T}{\beta},\\
&
\frac{\zeta_0}{\rho}\!=\!\frac{c_v(1+\alpha)^2}{a_v}\,\frac{v^2_{Tp}}{\beta n_\rho}\,
\mathcal{F}(\iota,\alpha)
\!=\!\frac{\varphi}{2}\frac{(1+\alpha)^2}{(1+\varphi)^2}\frac{\iota^2}{\tilde c_v^2}\frac{v_{Tp}^2}{\beta n_\rho},\\
&a_v\equiv2c_v+(1-\alpha)\alpha\,[(c_v+\iota)^2+c_v].
\end{align}
Here the ionization rate $\beta(T)$ is given by the approximate formula
Eq.~(\ref{ionization_rate})
It is remarkable that all parameters for sound propagation
in partially ionized hydrogen plasma are already \textit{ab-initio} calculated.

\begin{figure}[h]%
\includegraphics[scale=0.5]{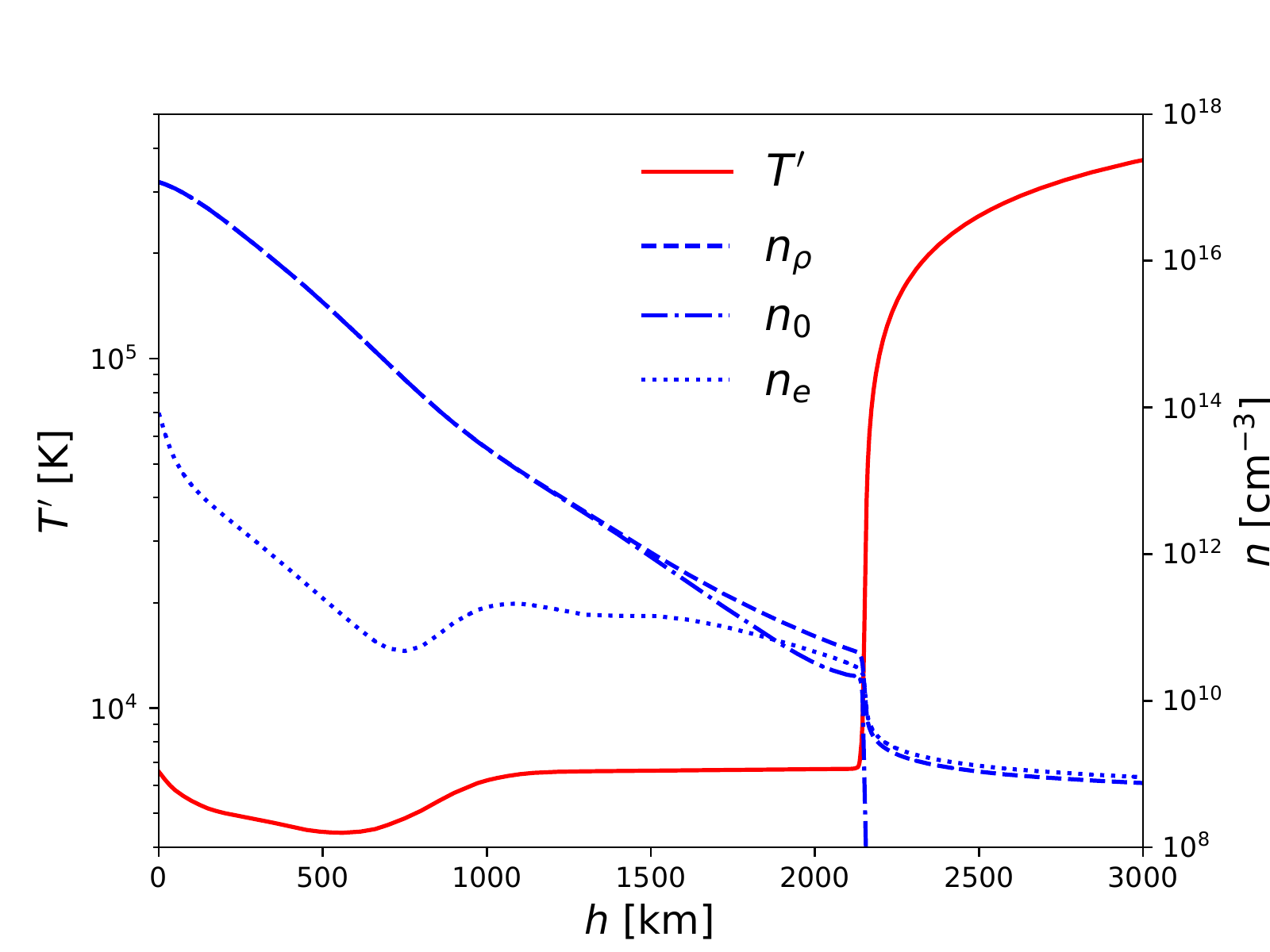}
\caption{Model C7 from Ref.~\onlinecite{Avret_Loeser}.
There are given height $h$ dependence of temperature $T^\prime$, mass density $n_\rho=n_p+n_0$,
density of electrons $n_e=n_p$ 
and density of neutral atoms $n_0$. 
}
\label{fig:AL-Prof}
\end{figure}

\begin{figure}[h]%
\includegraphics[scale=0.5]{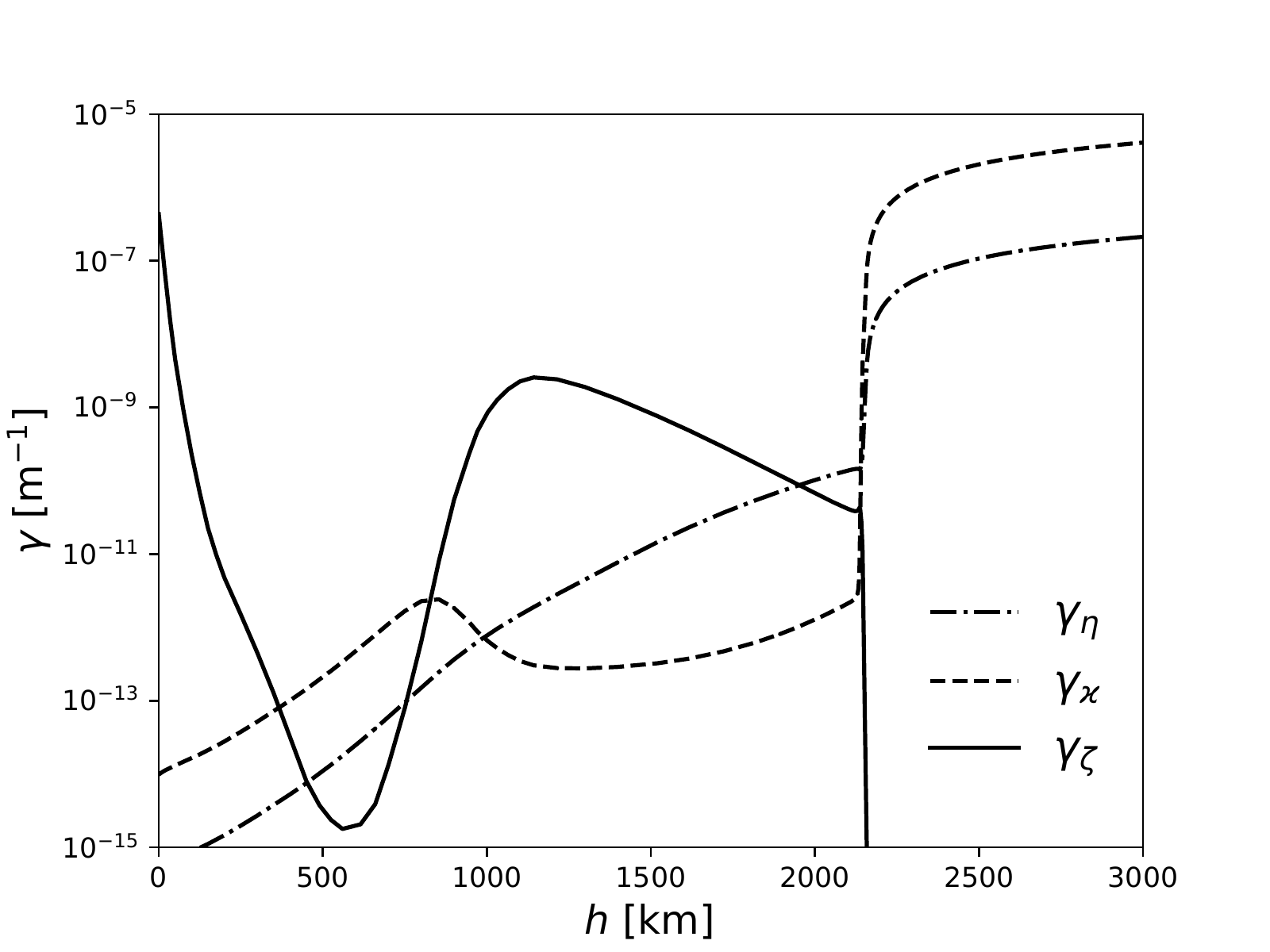}
\caption{Extinctions in the formula for the energy flux Eq.( \ref{total_extinction})
for frequency $f=100\,$mHz.
The compression viscosity term $\gamma_\zeta$ 
(the continuous line) dominates at small height and 
in the region with intensive ionization
which is one of the result of the present paper.
Due to small Prandtl number $\gamma_\varkappa>\gamma_\eta$
except the central region with intensive ionization 
when we have big heat capacity.
Roughly speaking the extinction coefficients 
are proportional to the heating power by wave damping.
}.
\label{fig:Gamma}
\end{figure}

\begin{figure}[h]%
\includegraphics[scale=0.5]{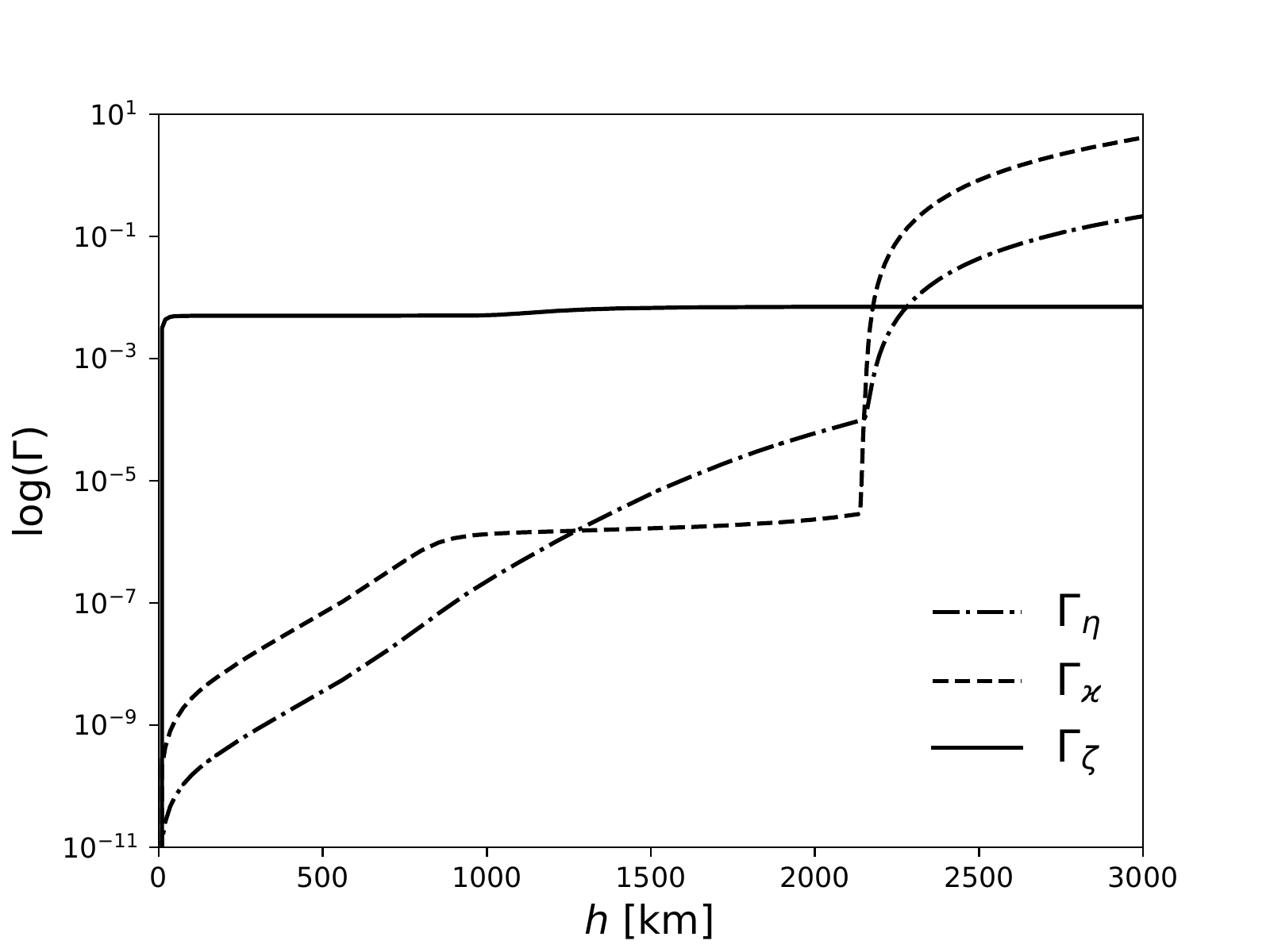}
\caption{Exponents in the formula for the energy flux Eq.~(\ref{Gamma})
One can see that bulk viscosity $\Gamma_\zeta$ dominates in the total absorption for height smaller than 2~Mm.
}.
\label{fig:Trs}
\end{figure}

In order to illustrate our results for the compression viscosity 
we use the model C7 by Avrett and Loeser\cite{Avret_Loeser}
for height $h$ dependence of the temperature $T$, 
hydrogen mass density $n_\rho$ and
electron density $n_e$ of the solar atmosphere
reproduced in \Fref{fig:AL-Prof}.
%%%%%%%%%%%%%%%%%%%%%%%%%%%
%%%%%%%%%%%%%%%%%%%%%%%%%%%
Having those profiles we can easily calculate 
different parts of the extinction as function of height
$\gamma_\eta$, $\gamma_\zeta$ and $\gamma_\varkappa$
depicted in \Fref{fig:Gamma} for frequency $f=100$~mHz
%%%%%%%%%%%%%%%%%%%%%%%%%%%5
%%%%%%%%%%%%%%%%%%%%%%%%%%%5
and corresponding quasi-classical exponents
according quasi-classical for the energy flux
Eq.~(\ref{energy flux}). 
\begin{align}&
\frac{q_s(h)}{q_s(0)}=\exp(-\Gamma)\approx 1-\Gamma,\quad
\Gamma=\Gamma_\eta+\Gamma_\zeta+\Gamma_\varkappa 
\label{Gamma}
,\\&
\Gamma_\eta\!=\!\int\limits_0^h\!2 \gamma_\eta\,\mathrm{d}x,\quad
\Gamma_\zeta\!=\!\int\limits_0^h\! 2\gamma_\zeta\,\mathrm{d}x,\quad
\Gamma_\varkappa\!=\!\int\limits_0^h\! 2\gamma_\varkappa\,\mathrm{d}x,
\end{align}
represented in \Fref{fig:Trs}.
As $\Gamma\ll1$ sound waves coming from the photo-sphere penetrates trough 
chromo-sphere and in this sense chromo-sphere is acoustically transparent.
However as $\Gamma_\zeta>\Gamma_\eta+\Gamma_\varkappa$ 
for $h<$2~Mm one can say that main channel of wave heating of the cromo-sphere
is trough compression viscosity created by ionization-recombination processes.
%%%%%%%%%%%
\begin{figure}[h]
\includegraphics[scale=0.5]{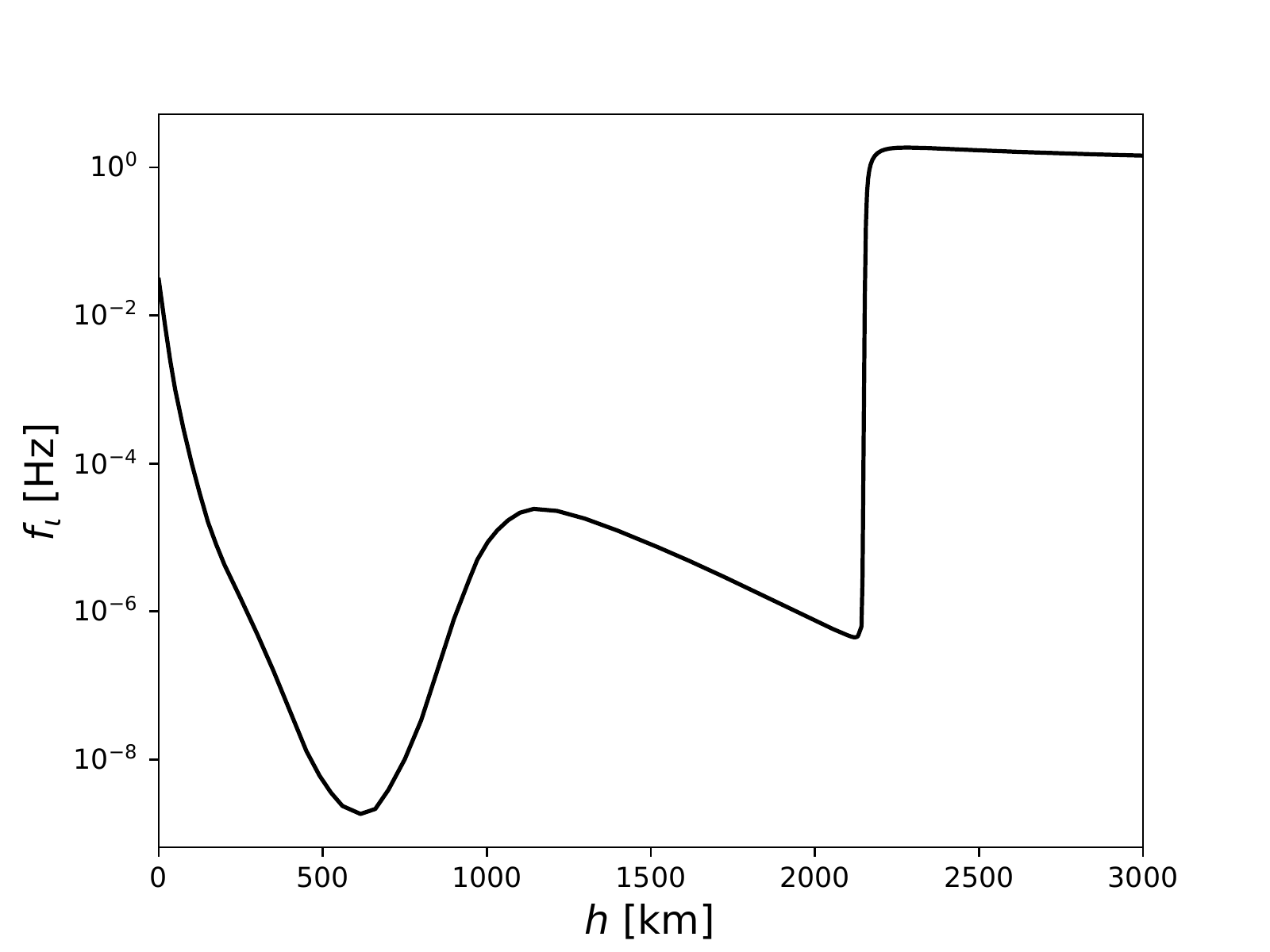}
\caption{Critical frequency $f_\iota(h)$ Eq.~(\ref{tau_ML}) 
above which the extinction 
$\gamma_\zeta$ is frequency independent.
One can see that even mHz waves are high frequent.
}
\label{fig:Freq}
\end{figure}        %
%%%%%%%%%%%
The height dependence of the critical frequency $f_\iota$ is drawn in \Fref{fig:Freq}
Additionally the corresponding dimensionless heat capacities per particle
$\tilde c_p$ and $\tilde c_v$ are represented in \Fref{fig:Cap}.
%%%%%%%%%%%%
\begin{figure}[h]   %
\includegraphics[scale=0.5]{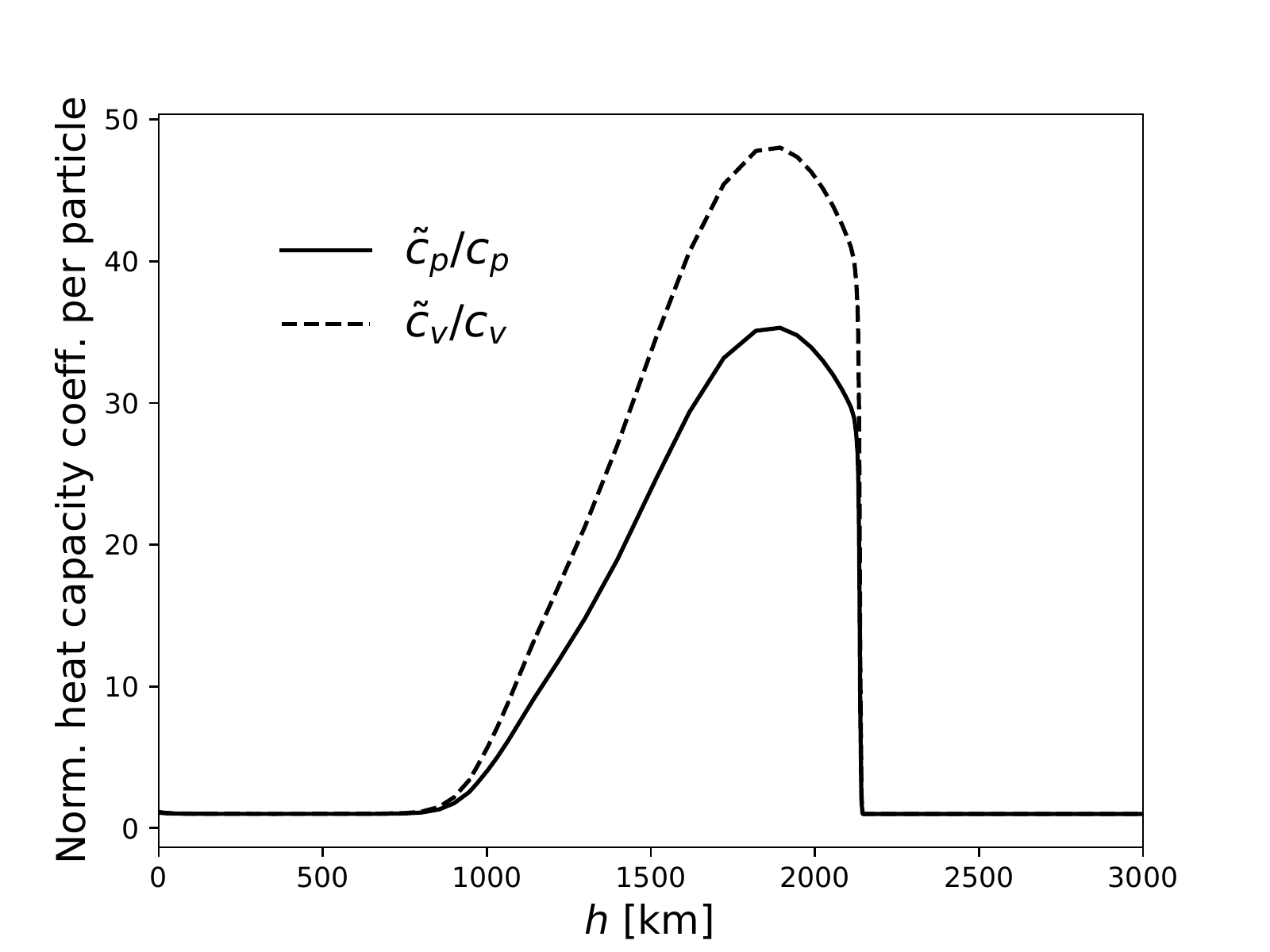}
\caption{Heat capacities per particle divided by its atomic values
according Eq.~\ref{c_p} and Eq.~\ref{c_v}. 
The significant re-normalization reveals that heat capacity below
2~Mm is related to ionization. 
The significant heat capacity ensures the 
relatively small variations of the temperature in the chromosphere.
}
\label{fig:Cap}
\end{figure}           %
%%%%%%%%%%%%

%%%%%%%%%%%%%%%%%%%%%%%%%%%%%%%
\section{Compression viscosity in other physical conditions}
It is curious to check whether 
the height temperature profile in the
partially ionized solar chromosphere 
can be described by absorption of the longitudinal 
sound waves coming from the
turbulent convection zone.
We suppose that frequencies much higher than 
Saha ionization relaxation rate $1/\tau$ give the main contribution.
No doubts the launch of the Parker Solar Probe satellite 
and its G\$ financing gives a new impetus on solar physics 
and physics of plasmas in general. 
And here we present a problem of H plasma which 
can be extended in many directions.

\textcolor{black}
{For sound absorption in different physical conditions Landau and Lifshitz consider
media with high thermal conductivity, moisture of two compounds and influence of the diffusion on the sound damping.\cite{LL6} Many other examples are considered in the monograph by Rudenko and Soluyan.\cite{Rudenko}
Also there are also many examples in the astrophysics, see for instance Ref.~\onlinecite{Zavershinskii}
and references therein.
That is why we consider as useful to mention also some other analogous application
of the theory we develop here.
From hydrodynamic point of view the microscopic mechanism 
of creation of the second viscosity is irrelevant.}

It is also interesting to mention that the damping of mechanical oscillations by chemical or ionization reactions can be observed in many common situations.
Pour beer or champagne 
into identical glasses and compare their deaf sound  
with the ringing of the of glasses full of wine or cognac produced by hitting them with a teaspoon.
What's the difference? Carbon dioxide in small bubbles.

For those whom some Lord forbids $\mathrm{C_2H_5OH}$, 
let they fill their glass with water 
and add a tablespoon of English salt
(Epsom salt, epsomite $\mathrm{MgSO_4\cdot 7 H_2O}$);
cheers.
What is thee difference between English salt
and Himalayan salt?
Partial  salvation
\be
\alpha=\mathrm{\frac{[Mg^{2+}]}
{[MgSO_4]+[Mg^{2+}]}}
\ee
versus almost complete one
\be
\alpha=\mathrm{\frac{[Na^{+}]}
{[Na^{+}]+[NaCl]}}.
\ee
The deaf sound of a glass of water with English salt 
is related to the big rate $\nu$ of the reaction 
\be
\mathrm{MgSO}_4\longleftrightarrow
\mathrm{Mg}^{2+}+\mathrm{SO}_4^{2-}
\ee
and the significant value of
$(1-\alpha)\alpha<0.25.$
However, the physico-chemistry of water 
solutions has no the transparent simplicity of the hydrogen plasma
and the analogy with bulk viscosity damping is rather qualitative.
\cite{Karim:52,Military,Francois:82,Neighbors:17}

The last example of damping is related to ionization reactions give
quartz resonators. 
Q-factors of those resonators 
can be significantly decreased due to Al impurities 
and changes of the ionization charge in Al ion by the strength 
of piezo-vibrations.

After this short \textcolor{black}{discussion} of related topics 
let us return to the H plasma and repeat the results derived in the present article:
1)~using the Wannier ionization cross-section we calculate
the rate $\beta$ of electron impact ionization
Eq.~(\ref{ionization_rate})
of the neutral atom,
and 2) simultaneously as inverse process 
the rate of two electron recombination 
$\gamma$ of the proton 
Eq.~(\ref{2e_recombination}).
3)~The entropy production $\dot S$ 
Eq.~(\ref{dot_S})
and the dissipation power $Q_\mathrm{ion}$
due to deviation of $\chi$ Eq.~(\ref{chi}) from the chemical equilibrium is a good basis for further consideration.
4)~Our main result is the damping rate of the sound waves
by oscillations of ionization.
5)~We propose a scenario for heating mechanisms of the
solar chromosphere and the explicit formula can be 
applied for existing models of the solar corona.
Up to now the heating mechanism of the solar 
chromosphere is an open problem and the height dependence
of the temperature $T(h)$ has not been calculated by 
first principle hydrodynamic calculations.
6)~In small magnetic fields Alfv\'en waves (AW) and 
Slow Magnetosonic Waves (SMW) have common dispersion
$\omega=V_\mathrm{A}k|\cos (\theta)|$,
where $\rho V_\mathrm{A}^2=B^2/\mu_0$
and $\theta$ is the angle between the constant external magnetic field and the wave vector.
However, the damping rates can be completely different
due to the pressure and temperature oscillations accompanying SMW.
Recently wave damping of of the MHD waves with his angular and plasma
beta dependence was analyzed in great detail by \cite{Perelomova:20}
in the case of completely ionized plasma where $\zeta=0$. 
It will be interesting to extend this analysis for the partially ionized hydrogen plasma.
We expect 
$\gamma_\mathrm{SMW}\gg \gamma_\mathrm{AW}$.
7)~The strong absorption of SMW in partially ionized plasmas
explains why magneto-hydrodynamic oscillations of the magnetic field are mainly orthogonal to the static magnetic field which is in qualitative agreement with the observations 
by different space missions. 
\textcolor{black}{It is time to put in the agenda of astrophysics
the problem of influence of di-atomic molecules dissociation on the second viscosity,
perhaps the first work in this direction was by Einstein.\cite{Einstein:20}}

Returning back to the hydrogen plasma we conclude that for the
first time some the second viscosity is calculated from first principles.

\acknowledgements

The authors would like to express their gratitude to
% Hassan Chamati,
Radostina Kamburova, Miroslav Georgiev, 
Iavor Boradjiev and Aleksander Petkov
for their interest in our work
and Tom Van Doorsselaere for pointing out 
Ref.~\onlinecite{Burgess}. 
The authors appreciate stimulating correspondence 
with Anna Perelomova
related to application of bulk viscosity in magneto-hydrodynamics.

\clearpage


\begin{thebibliography}{99}

\bibitem{Wannier:53}
G.~Wannier,
``The Threshold Law for Single Ionization of Atoms or Ions by Electrons'',
Phys. Rev. \textbf{90}(5), 817 (1953).

\bibitem{Fite:58}
W.~L.~Fite and R.~T.~Brackmann,
``Collisions of  electrons with Hydrogen Atoms. I. Ionization'',
Phys.~Rev. \textbf{112}(4), 1141%-1151
(1958).

\bibitem{Rothe:62}
E.~W. Rothe, L.~L.~Marino, R.~H.~Neynaber, and S.~M.~Trujillo,
``Electron Impact Ionization of Atomic Hydrogen and Atomic Oxygen'',
Phys. Rev. \textbf{125}, 582 (1962).

\bibitem{McGowan:67}
J.~W.~ McGowan, E.~M.~Clarke,
``Ionization of H(1s) near Threshold'',
Phys. Rev. \textbf{167}(1), 47 %-51
(1967).

\bibitem{Shyn:92}
T.~W.~Shyn,
``Doubly difFerential cross sections of secondary electrons ejected from atomic hydrogen by electron impact'',
Phys. Rev. A \textbf{47}, 2951 (1992).

\bibitem{Shakhatov:18}
V.~A.~Shakhatov and Yu.~A.~Lebedev,
``Analysis of Data on the Cross Sections for Electron-Impact Ionization and Excitation of Electronic States
of Atomic Hydrogen (Review)'',
Plas. Phys  Rep. \textbf{44}(1), 161%170
(2018).

\bibitem{LL3} 
L.~D.~Landau and E.~M.~Lifshitz, with participation of L.~P.~Pitaevskii,
\textit{Quantum mechanics. Non-relativistic theory} in
L.~D.~Landau and E.~M.~Lifshitz,
\textit{Landau-Lifshitz course on theoretical physics, Vol.~III}
(4th ed., Pergamon, New York, 1990),
Sec.~144, ``Scattering matrix at presence of reactions'', Eqs.~(144.15-16);
Sec.~147, ``Behavior of cross-section near threshold of reaction'', underline comment. 

\bibitem{LL4}
V.~B.~Berestetskii, E.~M.~Lifshitz and L.~P.~Pitaevskii,
\textit{Quantum Electrodynamics} in
L.~D.~Landau and E.~M.~Lifshitz,
\textit{Landau-Lifshitz Course on Theoretical Physics, Vol.~IV}
(2nd ed., Pergamon, New York, 1981),
Sec.~56, ``Photo-effect. Non-relativistic case'',
Eq.~(56.13) and Eq.~(56.15).

\bibitem{Pitaevskii:62}
L.~P.~Pitaevskii, ``Electron Recombination in a Monatomic gas'', JETP \textbf{15}(5), 919-921 (1962), Eq.~(19).

\bibitem{LL10}
E.~M.~Lifshitz and L.~P.~Pitaevskii,
\textit{Physical Kinetics} in
L.~D.~Landau and E.~M.~Lifshitz,
\textit{Landau-Lifshitz Course on Theoretical Physics, Vol.~X}
(Pergamon, New York, 2002), Sec.~2, ``Detailed balance principle'', 
Eq.~(2.9) and Eq.~(3.7), Sec.~24, ``Recombination and ionization'', Problems 1 and~2,
Sec. 43 ``Mean free path of particles in plasma'', Eqs.~(43.-9-10).

\bibitem{BelyaevBudker:58}
S.~T.~Belyaev and G.~I.~Budker,
``Multi-quantum Recombination in an Ionised Gases'',
in \textit{Fizika Plazmy i Problemy Upravlyayemykh Termoyadernykh Reaktsii},
ed. by M.~A.~Leontovich (USSR Academy of Science, Moscow and Leningrad, 1958), Vol.~III, 41-49 (in Russian);

\textit{Plasma Physics and the Problem of Controlled Thermonuclear Reactions},
ed. by M.~A.~Leontovich (Pergamon Press, London, 1959), Vol.~III, 45-55.

\bibitem{Hill}
T.~L.~Hill, 
\textit{An Introduction to Statistical Thermodynamics}
(Addison-Wesley, Massachusetts, 1960), 
Chapter 10 ``Chemical Equilibrium in Ideal Gas Mixtures", Eqs.~(10.21-23).

\bibitem{KittelCramer}
C.~Kittel and H.~Cramer,
\emph{Statistical Thermodynamics} (2 ed., W.~H.~Freeman and Company, San Francisco, 1980).

\bibitem{LL5}
L.~D.~Landau, E.~M.~Lifshitz and L.~P.~Pitaevskii,
\textit{Statistical Physics Part 1} in
L.~D.~Landau and E.~M.~Lifshitz,
\textit{Landau-Lifshitz Course on Theoretical Physics, Vol.~V}
(3rd ed., Pergamon Press, New York, 1980), 
Sec.~43, ``Ideal gas with constant heat capacity'';
Sec.~45, ``Mono-atomic gas'';
Sec.~46, ``Mono-atomic gas. Influence of electronic momentum'', Eq.~(46.1a);
Sec.~78, ``Thermodynamic variables of classical plasma'', Eq.~(78.8);
Sec.~102, ``The law of mass action'', Eqs.~(102.4) and (102.8);
Sec.~103, ``Heat of reaction'', Eq.~(103.5);
Sec.~104, ``Ionization equilibrium'', Eq.~(104.2)
Chap.~XIV, ``Type-II Phase transitions and critical phenomena'',
Eqna.~(143.1-6).

\bibitem{Saha}
%M.~N.~Saha, ``LIII. Ionization in the solar chromosphere'',
%Phil. Mag. Ser. 6, \textbf{40}(238), 472-488 (1920);
%
M.~N.~Saha, ``On a physical theory of stellar spectra'', 
Proc. R. Soc. Lond. A, \textbf{99}(697), 135-153 (1921).

\bibitem{Mandelstam:37}
L.~I.~Mandelstam and M.~A.~Leontovitch,
J. Exp. Theor. Phys. \textbf{7}(3), 438 (1937), (in Russian);
M.~A.~Leontovitch,
J. Exp. Theor. Phys. \textbf{6} 561 (1936), (in Russian);
\\
M.~A.~Leontovich, \textit{Selected Works. Theoretical Physics.}
(Nauka, Moscow, 1985), (in Russian);\\
L.~I.~Mandelstam, \textit{Collected Works, vol. 2}, ed. S.~M.~Rytov
(Academy of Sciences of the Soviet Union, 1947), p.~176, (in Russian).

\bibitem{LL6}
L.~D.~Landau and E.~M.~Lifshitz,
\emph{Course on Theoretical Physics, Vol.~6, Fluid Mechanics}
(2nd ed., Pergamon Press, New York, 1987), 
Sec.~1 ``Continuity equation'', ``liquid particle'',
Sec.~25 ``Attenuation of gravitational waves'',
Eq.~(50.3) and Eq.~(53.4), 
Sec.~64 ``Energy and momentum of sound waves'',
Sec.~79, ``Sound absorption'', Eqs.~(79.3-6),
Sec.~78, ``Second viscosity'', Eq.~(78.10), 
according 3-rd edition Eqs.~(81.1-14)

\bibitem{DeGroot} S.~R.~De~Groot and P. Masur,
\textit{Non-Equilibrium Thermodynamics},
(Dover Publications, New York, 1974); Chap.~XII
``Viscous flow and relaxation phenomena'', 
Sec.~``Acoustic relaxation'', 
See also in the subject index ``viscosity''. 

\bibitem{LL1}
L.~D.~Landau and E.~M.~Lifshitz,
\emph{Course on Theoretical Physics, Vol.~1, Mechanics}
(4th ed., Pergamon Press, New York, 1989), 
Sec.~51, ``Accuracy of conservation of adiabatic invariant'', Problem 2.

\bibitem{LL8}         
L.~D.~Landau and E.~M.~Lifshitz,	
\textit{Electrodynamics in Continuous Media} in 
L.~D.~Landau and E.~M.~Lifshitz,	
\textit{Course of Theoretical Physics, Vol.~VIII}
(Pergamon Press, New York, 1960), Sec.~69 ``Hydromagnetic waves'',
Problem: ``Damping of Alfv\'en waves.''

\bibitem{gamma_hydrogen}
T.~M.~Mishonov, I.~M.~Dimitrova, and A.~M.~Varonov,
``On the Influence of the Ionization-Recombination Processes on Hydrogen Plasma Polytropic'', arXiv:2008.03565.

\bibitem{Newton}
I.~Newton,
\textit{Philosophiæ Naturalis Principia Mathematica}
(5 July 1686); \url{https://www.wdl.org/en/item/17842/view/1/3/}.
%\url{https://en.wikipedia.org/wiki/Philosophi%C3%A6_Naturalis_Principia_Mathematica}.

\bibitem{Rudenko}
O.~V.~Rudenko, S.~I.~Soluyan
"Theoretical foundations of nonlinear acoustics", in
\textit{Studies in Soviet Science}, ed. O.~V.~Rudenko (Springer, Berlin, 1977);
Section IV. ``On the dispersion properties of media. Medium with relaxation'',
Eqs.~(IV.1.1-26); translated from Russian original by R.~T.~Beyer;\\
(Consultant Bureau, New York, Plenum, 1977).

\bibitem{Burgess}
A.~ Burgess and M.~J.~ Seaton,
``The ionization equilibrium for iron in the solar corona'',
Monthly Notices of the Royal Astronomical Society (MNRAS),
\textbf{127}, 355-%358
(1964); DOI 10.1093/mnras/127.5.355 

\bibitem{Zavershinskii} D.~I.~Zavershinskii, D.~Y.~Kolotkov, V.~M.~Nakariakov,
N.~E.~Molevich, and D.~S.~Ryashnikov,
``Formation of quasi-periodic slow magnetoacoustic wave trains by heating/cooling misbalance''
Phys. Plasmas \textbf{26}, 082113 (2019).

\bibitem{Karim:52}
S.~M.~Karim and L.~Rosenhead,
``The Second Coefficient of Viscosity of Liquids and Gases'',
Rev. Mod. Phys. \textbf{24}(2), 108 (1952).

\bibitem{Military}
R.~H.~Mellem, P.~M.~Scheifele and D.~G.~Browning,
``Global Model for Sound Absorption in Sea Water'',
NUSC Techn. Report 7923, AD-A181 688 (1937); Reviewed and Approved: 14 May 1987.

\bibitem{Francois:82}
R.~E.~Francois and G.~R.~Garrison,
``Sound absorption based on ocean measurements: Part I: Pure water and magnesium sulfate contributions'',
J. Acoust. Soc. Am. \textbf{72}(3), 896 (1982).

\bibitem{Neighbors:17}
T.~H.~Neighbors, III,
``Chapter 4 - Absorption of Sound in Seawater'', 
in Leif Bj\o rn\o, \textit{Applied Underwater Acoustics},
eds. T.~H.~Neighbors, III and D.~Bradley (Elsevier, Amsterdam, Netherlands, 2017).

\bibitem{Avret_Loeser} 
E.~H.~Avrett and R.~Loeser, 
``Models of the Solar Chromosphere and Transition Region from SUMER and HRTS Observations: Formation of the Extreme-Ultrablack Spectrum of Hydrogen, Carbon and Oxygen'', Astrophys. J. Supp. Ser. \textbf{175}, 229--276 (2008), Fig.~8 and Table~26.

\bibitem{Perelomova:20}
A.~Perelomova,
``On description of periodic magnetosonic perturbations in a quasi-isentropic plasma with
mechanical and thermal losses and electrical resistivity''
Phys. Plasmas \textbf{27}, 032110 (2020).
%; doi: 10.1063/1.5142608

\bibitem{Einstein:20} A.~Einstein, Sitzungsber., Preussische Akad. Wiss., Berlin (1920) 380.

\end{thebibliography}
\end{document}